\documentclass[acmsmall,screen,rgb]{acmart}

\usepackage[ruled,linesnumbered,noend]{algorithm2e}
\usepackage{algorithmic}
\usepackage{bbm}
\usepackage{proof}
\usepackage{mathpartir}
\usepackage{wrapfig}

\usepackage{xparse}
\usepackage{enumitem}
\AtBeginDocument{%
}

\newcommand{\AC}{\ensuremath{\sf AC}}
\newcommand{\ACsp}{\ensuremath{\sf ACsp}}
\newcommand{\SL}{\ensuremath{\sf SL}}

\newcommand{\transrel}{\ensuremath{\Delta}} 
\newcommand{\controls}{\ensuremath{Q}} 
\newcommand{\defn}[1]{\textit{#1}} 
\newcommand{\ialphabet}{\ensuremath{\Sigma}}
\newcommand{\conf}{\alpha}

\newcommand{\Free}[1]{\ensuremath{\text{free}(#1)}}
\newcommand{\Var}[1]{\ensuremath{\text{var}(#1)}}

\newcommand{\OMIT}[1]{}

\newcommand{\weight}[1]{W(#1)}

\SetCommentSty{mycommfont}
\newlength{\commentWidth}
\newif\ifdraft\drafttrue
\ifdraft
\newcommand\anthony[1]{{\color{blue}
[#1 - \textbf{Anthony}]}}
\newcommand\philipp[1]{{\color{magenta}
[#1 - \textbf{Philipp}]}}
\newcommand\lukas[1]{{\color{teal}
[#1 - \textbf{Lukas}]}}
\newcommand\tomas[1]{{\color{orange}
[#1 - \textbf{Tomas}]}}
\else
\newcommand\tomas[1]{}
\newcommand\lukas[1]{}
\newcommand\anthony[1]{}
\newcommand\philipp[1]{}
\fi

\newcommand\shortlong[2]{#2}

\usepackage{color}
\shortlong{}{\pagestyle{plain}}

\DeclareRobustCommand{\myEnsuremath}{%
  \ifmmode
    \expandafter\@firstofone
  \else
    \expandafter\@myEnsuredmath
  \fi}
\long\def\@myEnsuredmath#1{\m{\relax#1}}

\newcommand{\concat}{\circ}
\newcommand{\init}{I}
\newcommand{\final}{F}
\newcommand{\afa}{\mathcal{A}}
\newcommand{\aft}{\mathcal{R}}

\newcommand{\bofo}[1]{\mathbb{F}_{\!#1}}
\newcommand{\pobofo}[1]{\mathbb{F}_{\!#1}^+}
\newcommand{\nebofo}[1]{\mathbb{F}_{\!#1}^-}

\newcommand{\states}{Q}
\newcommand{\transitions}{\Delta}
\newcommand{\bool}{\mathbb{B}}
\newcommand{\symb}{b}
\newcommand{\nat}{\mathbb{N}}
\newcommand{\Z}{\mathbb{Z}}
\newcommand{\power}[1]{\mathcal{P}(#1)}

\newcommand{\trel}{\mathit{Trans}}

\newcommand\aftrel{\mathcal{R}}
\newcommand\aftrels{\mathcal{S}}
\newcommand\afare{P}
\newcommand\rel{\mathit{Rel}}

\newcommand\sass{\iota}
\newcommand\pass{\nu}
\newcommand{\run}{\rho}
\renewcommand{\conf}{\run}
\newcommand{\vect}[1]{\bar{#1}}
\DeclareDocumentCommand\ap{ g }{%
    {\mathit{A\!P}%
        \IfNoValueF {#1} {\times [#1]}%
    }%
}

\newcommand{\bvar}{V}
\newcommand{\bvarn}{V_n}
\newcommand{\bv}{b}
\newcommand{\BV}{\power{\bvar}}

\newcommand{\bvare}{W}
\newcommand{\bvarex}[1]{\bvare\langle{#1}\rangle}
\newcommand{\bvarek}{\bvarex{k}}
\newcommand{\bvarxx}[2]{{#1}\langle{#2}\rangle}

\newcommand{\compact}[1]{\rangle{#1}\langle}
\newcommand{\select}[2]{{#1}\hspace*{-1mm}\downarrow_{#2}}

\DeclareDocumentCommand\bvarn{ g }{%
    {V_n%
        \IfNoValueF {#1} {\times [#1]}%
    }%
}

\newcommand{\true}{\mathtt{true}}
\newcommand{\false}{\mathtt{false}}

\newcommand{\polarity}[1]{\widetilde{#1}}

\newcommand{\eps}{E}
\newcommand{\epsuni}{\mathcal{E}}

\DeclareDocumentCommand\bvarne{ g }{%
    {(\bvarn \cup \eps)%
        \IfNoValueF {#1} {\times [#1]}%
    }%
}

\newcommand{\letterof}[2]{{#1}[{#2}]}
\newcommand{\nicemod}{\mathtt{~mod~}}

\newcommand{\strsolver}{\textsc{Sloth}}
\newcommand{\slog}{\textsc{SLOG}}
\newcommand{\cvc}{\textsc{CVC4}}
\newcommand{\stranger}{\textsc{Stranger}}
\newcommand{\sthreep}{\textsc{S3P}}
\newcommand{\smtlib}{\textsc{SMT-LIBv2}}

\newcommand{\ASSERT}[1]{\textbf{assert}(#1)}
\newcommand{\SKIP}{\textbf{skip}}

\usepackage{booktabs}   
\usepackage{subcaption} 

\setcopyright{none}
\acmJournal{PACMPL}
\acmYear{2018} 
\copyrightyear{2018}
\acmVolume{2} \acmNumber{POPL} \acmArticle{4} \acmMonth{1} \acmPrice{}\acmDOI{10.1145/3158092}

\begin{document}

\title[String Constraints with Concatenation and Transducers Solved Efficiently]{String Constraints with Concatenation and Transducers Solved
Efficiently \shortlong{}{(Technical Report)}}


\author{Luk\'{a}\v{s} Hol\'{i}k}
\orcid{0000-0001-6957-1651}             
\affiliation{
  \institution{Brno University of Technology}    
  \department{Faculty of Information Technology, IT4Innovations Centre of Excellence} 
  \streetaddress{Bo\v{z}et\v{e}chova 2}
  \city{Brno}
  \postcode{CZ-61266}
  \country{Czech Republic}
}
\email{holik@fit.vutbr.cz}          

\author{Petr Jank{\r u}}
\orcid{nnnn-nnnn-nnnn-nnnn}             
\affiliation{
  \institution{Brno University of Technology}    
  \department{Faculty of Information Technology, IT4Innovations Centre of Excellence} 
  \streetaddress{Bo\v{z}et\v{e}chova 2}
  \city{Brno}
  \postcode{CZ-61266}
  \country{Czech Republic}
}
\email{ijanku@fit.vutbr.cz}         

\author{Anthony W. Lin}
\orcid{0000-0003-4715-5096}
\affiliation{
  \department{Department of Computer Science}              
  \institution{University of Oxford}            
  \streetaddress{Wolfson Building, Parks Road}
  \city{Oxford}
  \postcode{OX1 3QD}
  \country{United Kingdom}
}
\email{anthony.lin@cs.ox.ac.uk}

\author{Philipp R\"ummer}
\orcid{0000-0002-2733-7098}             
\affiliation{
  \department{Department of Information Technology}             
  \institution{Uppsala University}           
  \streetaddress{Box 337}
  \city{Uppsala}
  \postcode{75105}
  \country{Sweden}
}
\email{philipp.ruemmer@it.uu.se}         

\author{Tom\'{a}\v{s} Vojnar}
\orcid{0000-0002-2746-8792}             
\affiliation{
  \institution{Brno University of Technology}    
  \department{Faculty of Information Technology, IT4Innovations Centre of Excellence} 
  \streetaddress{Bo\v{z}et\v{e}chova 2}
  \city{Brno}
  \postcode{CZ-61266}
  \country{Czech Republic}
}
\email{vojnar@fit.vutbr.cz}         


\begin{abstract}
    String analysis is the problem of reasoning about how strings are 
    manipulated by a program. It has numerous applications including
    automatic detection of cross-site scripting, and automatic test-case 
    generation. A~popular string analysis technique includes symbolic
    executions, which at their core use constraint solvers over the string
     domain, a.k.a.\ string solvers.
     Such solvers
     typically reason about constraints expressed in theories over strings with 
     the concatenation operator
     as an atomic constraint. In recent years, researchers started to recognise
     the importance of incorporating the replace-all operator (i.e. replace
     all occurrences of a string by another string) and, more generally,
     finite-state transductions in the theories of strings with concatenation. 
     Such string operations are typically crucial for reasoning
     about XSS vulnerabilities in web applications, especially for modelling 
     sanitisation functions and implicit browser transductions (e.g. innerHTML). 
     Although this
     results in an undecidable theory in general, it was recently shown that
     the straight-line fragment of the theory is decidable, and is sufficiently
     expressive in practice.
     In this paper, we provide the first string solver that
     can reason about constraints involving both concatenation and
     finite-state transductions. Moreover, it has a completeness and termination
     guarantee for several important fragments (e.g. straight-line fragment).
     The main challenge addressed
    in the paper is the prohibitive worst-case complexity of the 
    theory (double-exponential time), which is exponentially harder than the
    case without finite-state transductions. To this end, we propose a method
    that exploits succinct alternating finite-state automata as concise
    symbolic representations of string constraints.
    In contrast to previous approaches using nondeterministic automata,
    alternation offers not only 
    exponential savings in space when representing Boolean
    combinations of transducers, but also a possibility
    of succinct representation of otherwise costly combinations of
    transducers and concatenation.
    Reasoning about the emptiness of the AFA language requires
    a state-space exploration in an exponential-sized graph, for which
    we use model checking algorithms (e.g.~IC3). We have implemented our
    algorithm and demonstrated its efficacy on benchmarks that are
    derived from cross-site scripting analysis
    and  other  examples in the literature. 
\end{abstract}

\ccsdesc[500]{Theory of computation~Automated reasoning}
\ccsdesc[500]{Theory of computation~Verification by model checking}
\ccsdesc[500]{Theory of computation~Program verification}
\ccsdesc[500]{Theory of computation~Program analysis}
\ccsdesc[300]{Theory of computation~Logic and verification}
\ccsdesc[100]{Theory of computation~Complexity classes}


\keywords{String Solving, Alternating Finite Automata, Decision Procedure,
IC3}  

\maketitle

\section{Introduction}

Strings are a fundamental data type in many programming languages. This
statement is true now more than ever, especially owing to the rapidly growing 
popularity of
scripting languages (e.g. JavaScript, Python, PHP, and Ruby) wherein
programmers tend to make heavy use of string variables. String manipulations
are often difficult to reason about automatically, and could easily lead to 
unexpected programming errors. In some applications, some of these errors could 
have serious security
consequences, e.g., cross-site scripting (a.k.a. XSS), which are ranked among
the top three classes of web application security vulnerabilities by OWASP
\cite{OWASP}. 

Popular methods for analysing how strings are being manipulated by a 
program include \emph{symbolic executions}
\cite{Berkeley-JavaScript,BTV09,RVG12,KS16,cadar-icse11,jalangi,EXE,DART,ExpoSE} which 
at their core use 
constraint solvers over the string domain (a.k.a.\ \emph{string solvers}). 
String solvers have been the subject of numerous papers in the past decade,
e.g., see 
\cite{Stranger,fmsd14,Berkeley-JavaScript,S3,BTV09,Abdulla14,cvc4,HW12,Yu09,Balzarotti08,symbolic-transducer,ganesh-word,Was08a,FL10,FPBL13,BEK,HAMPI,Z3-str,yu2011,DV13,LB16,joxan-cav16,tinelli-fmsd16,tinelli-hotsos16,tinelli-frocos16}
among many others. As is common in constraint solving, we follow
the standard approach of \emph{Satisfiability Modulo Theories (SMT)} 
\cite{SMT-CACM}, which is an extension of the problem of satisfiability of
Boolean formulae wherein each atomic proposition can be interpreted over some
logical theories (typically, quantifier-free). 

Unlike the case 
of constraints over integer/real arithmetic (where many decidability and 
undecidability results are known and powerful algorithms are
already available, e.g., the simplex algorithm), string constraints are much
less understood. This is because there are many different string operations
that can be included in a theory of strings, e.g., concatenation,
length comparisons, regular constraints (matching against a regular 
expression), and replace-all (i.e. replacing every occurrence of a string
by another string). Even for the theory of strings with the
concatenation operation alone, existing string solver cannot
handle the theory (in its full generality) in a sound and complete manner,
despite the existence of a theoretical decision procedure for the problem
\cite{Gut98,Makanin,Plandowski,Plandowski06,jez16,diekert}. This situation
is exacerbated when we add extra operations like string-length comparisons,
in which case even decidability is a long-standing open problem
\cite{ganesh-word}. 
In addition, recent works in string solving have argued in favour of adding the 
replace-all operator or, more generally finite-state transducers, to string 
solvers \cite{LB16,joxan-cav16,fmsd14,Stranger} in view of their importance for
modelling relevant sanitisers (e.g. backslash-escape) and implicit browser 
transductions (e.g. an application of HTML-unescape by innerHTML), e.g., see
\cite{DV13,symbolic-transducer,BEK} and Example \ref{ex:cacm} below. 
However,
naively combining the replace-all operator and concatenation yields
undecidability \cite{LB16}.


\begin{example}
    \label{ex:cacm}
    The following JavaScript snippet---an adaptation of an 
    example from \cite{Kern14,LB16}---shows use of
    \emph{both} concatenation and finite-state transducers:
    \OMIT{
    Consider the following simple JavaScript code snippet adapted from 
    \cite{Kern14}:
}
{\footnotesize
\begin{verbatim}
var x = goog.string.htmlEscape(name);
var y = goog.string.escapeString(x);
nameElem.innerHTML = '<button onclick= "viewPerson(\'' + y + '\')">' + x + '</button>';
\end{verbatim}
}
The code assigns an HTML markup for a button to the DOM element 
    \texttt{nameElem}. Upon click, the 
    button will invoke the function \texttt{viewPerson} on the input 
    \texttt{name} whose value is an untrusted variable. The code attempts to 
    first sanitise the value of \texttt{name}.
This is done via The Closure Library \cite{closure} string functions 
\texttt{htmlEscape} 
and \texttt{escapeString}.
    \OMIT{
To check the above code for XSS vulnerabilities, one typically has a list of
\emph{attack patterns} (concrete strings or regular expressions) for 
dangerous lines in the program (those with functions that can
insert strings to the DOM, e.g., \texttt{innerHTML} or 
\texttt{document.write()}). One possible attack pattern for our example is the
string
}
Inputting the value \texttt{Tom \& Jerry} into \texttt{name} gives the desired
HTML markup:
{\footnotesize
    \begin{verbatim}
<button onclick="viewPerson('Tom &amp; Jerry')">Tom &amp; Jerry</button>
    \end{verbatim}
    }
On the other hand, inputting value {\ttfamily \verb+');attackScript();//+}
to \texttt{name}, results in the markup:
    {\footnotesize
    \begin{verbatim}
<button onclick="viewPerson('&#39;);attackScript();//')">&#39;);attackScript();//')</button>
    \end{verbatim}
    }
Before this string is inserted into the DOM via \texttt{innerHTML}, an 
implicit browser transduction will take place \cite{web-model,mXSS}, i.e., 
    HTML-unescaping the string inside the 
\texttt{onclick} attribute and then invoking the attacker's script 
\texttt{attackScript()} after \texttt{viewPerson}.
This subtle DOM-based XSS bug is due to calling the right escape
functions, but in wrong order. \qed
\end{example}


One theoretically sound approach proposed in \cite{LB16} for overcoming the undecidability of string
constraints with both concatenation and finite-state transducers is to impose a 
\emph{straight-line restriction} on the shape of constraints.  This
straight-line fragment can be construed as the problem of \emph{path
feasibility}
\cite{BTV09} in the following simple imperative language (with only
assignment, skip, and assert) for defining non-branching and non-looping 
string-manipulating programs that are generated by symbolic execution:
\begin{equation*}
    S ::= y := a \ |\ \text{\ASSERT{$b$}}\ |\ \text{\SKIP}\ |\ S_1; S_2, \qquad
    a ::= f(x_1,\ldots,x_n), \qquad b ::= g(x_1) 
\end{equation*}
where $f: (\ialphabet^*)^n \to \ialphabet^*$ is either an application of
concatenation
$x_1 \concat \cdots \concat x_n$ or an application of a~finite-state 
transduction $R(x_1)$, and
$g$ tests membership of $x_1$ in a regular language. Here, some variables are
undefined ``input variables''. Path feasibility
asks if there exist input strings that satisfy all assertions and applications
of transductions in the program. 
It was shown in \cite{LB16} that such a
path feasibility problem (equivalently, satisfiability for the aforementioned
straight-line fragment) is decidable. As noted
in \cite{LB16} such a fragment can express the program logic of many 
interesting examples of string-manipulating programs with/without XSS 
vulnerabilities. For instance, the above example can be modelled as a 
straight-line formula where the regular constraint comes from an attack 
pattern like the one below:
\OMIT{
\begin{verbatim}
<a onclick="viewPerson(''); attackScript(); [^'"]*"> [^'"]* </a>
\end{verbatim}
}
\begin{verbatim}
e1 = /<button onclick=
        "viewPerson\(' ( ' | [^']*[^'\\] ' ) \); [^']*[^'\\]' \)">.*<\/button>/
\end{verbatim}
Unfortunately, the decidability proof given in \cite{LB16} provides only
a theoretical argument for decidability and complexity upper bounds (an
exponential-time reduction to the \emph{acyclic fragment} of intersection
of rational relations\footnote{This fragment consists of constraints that are
given as conjunctions of transducers $\bigwedge_{i=1}^m R_i(x_i,y_i)$, wherein 
the graph $G$ of variables does not contain a cycle. The graph $G$ contains
vertices corresponding to variables $x_i,y_i$ and that two variables
$x, y$ are linked by an edge if $x = x_i$ and $y = y_i$ for some $i \in
\{1,\ldots,m\}$.} whose decidability proof in turn is a highly intricate 
polynomial-space procedure using Savitch's trick \cite{BFL13}) and does
not yield an implementable solution. Furthermore,
despite its decidability, the string logic has a prohibitively high complexity
(EXPSPACE-complete, i.e., exponentially higher than  without 
transducers), which could severely limit its applicability.

\paragraph{Contributions}
Our paper makes the following contributions to overcome the above
challenges:
\begin{enumerate}
\item We propose a fast reduction of satisfiability of formulae in the
  straight-line fragment and in the acyclic fragment to the emptiness
  problem of \emph{alternating finite-state automata (AFAs).}
  The reduction is in the worst case exponential in the number of concatenation
        operations\footnote{This is an unavoidable computational limit imposed
        by EXPSPACE-hardness of the problem \cite{LB16}.}, but otherwise 
        polynomial in the size of a formula.
%
  In combination with fast model checking algorithms (e.g.~IC3 \cite{Bra12})
  to decide AFA emptiness, this yields the first practical 
    algorithm for handling
  string constraints with concatenation, finite-state transducers
  (hence, also replace-all), and regular constraints, and a decision
  procedure for formulae within the straight-line and acyclic
  fragments.
\item We obtain a substantially simpler proof for the decidability and
  PSPACE-membership of the acyclic fragment of intersection of
  rational relations of \cite{BFL13}, which was crucially used in
  \cite{LB16} as a blackbox in their decidability proof of the
  straight-line fragment.
\item We define optimised translations from AFA emptiness to reachability over
  Boolean
  transition systems (i.e. which are succinctly represented
  by Boolean formulae). We implemented our algorithm for string constraints in 
        a new string solver called \strsolver, and provide an extensive 
        experimental evaluation. \strsolver\
  is the first solver that can handle string constraints that arise
  from HTML5 applications with sanitisation and implicit browser
  transductions. Our experiments suggest that the translation to AFAs
  can circumvent the EXPSPACE worst-case complexity of the
  straight-line fragment in many practical cases.
\end{enumerate}
\paragraph{An overview of the results}
The main technical contribution of our paper is a new method for
exploiting alternating automata (AFA) as a succinct symbolic representation for representing formulae
in a complex string logic admitting concatenation and finite-state 
transductions.
In particular, the satisfiability problem for the string logic is
reduced to AFA language emptiness, for which we exploit fast model 
checking algorithms.
%
Compared to previous methods \cite{LB16,Abdulla14} that are based on 
nondeterministic automata (NFA) and transducers, we show that AFA can incur
\emph{at most a linear blowup} for each string operation permitted in the logic 
(i.e. concatenation, transducers, and regular constraints).
While the product NFA representing the intersection 
of the languages of two automata $A_1$ and $A_2$ would be
of size $O(|A_1| \times |A_2|)$, 
the language can be represented using an AFA of size
$|A_1|+|A_2|$ (e.g. see \cite{Vardi-banff}). The difficult cases are how to
deal with concatenation and replace-all, which are our contributions to the
paper. More precisely, a constraint of the form $x := y.z \wedge x \in L$ (where $L$ is the language accepted by an automaton $A$) was 
reduced in
\cite{LB16,Abdulla14} to regular constraints on $y$ and $z$ by means of 
splitting $A$, which causes a cubic blow-up (since
an ``intermediate state'' in $A$ has to be guessed,
and for each state a product of two
automata has to be constructed). Similarly, taking the post-image
$R(L)$ of $L$ under a relation $R$ represented by a finite-state transducer $T$ gives us
an automaton of size $O(|T| \times |A|)$. 
%
A na\"ive application of AFAs is not helpful for those cases, since also
projections on AFAs are computationally hard.

The key idea to overcome these difficulties is to \emph{avoid} 
applying projections altogether, and instead use the AFA to represent general 
$k$-ary \emph{rational relations} (a.k.a. $k$-track finite-state transductions
\cite{Berstel,Sakarovitch,BFL13}).
This is possible because we focus on formulae without negation,
so that the (implicit) existential quantifications for 
applications of transducers can be placed outside the constraint. This means
that our AFAs operate on alphabets that are exponential in size
(for $k$-ary relations, the alphabet is $\{\epsilon,0,1\}^k$).
%
To address this problem, we introduce a succinct flavour of AFA with 
symbolically represented transitions. 
Our definition 
is similar to the concept of alternating symbolic automata in 
\cite{dantoni:alternating} with one difference.
While symbolic AFA take a transition $q \rightarrow_\psi \varphi$ from a state 
$q$ to a set of states satisfying a formula $\varphi$ if the input symbol 
satisfies a formula $\psi$,
our succinct AFA can mix constraints on successor states with those on 
input symbols within a single transition formula 
(similarly to the symbolic transition representation of deterministic automata in MONA \cite{monasecrets}, where sets of transitions are represented as multi-terminal BDDs with states as terminal nodes).
We show how automata splitting can be achieved with at most
linear blow-up. 
\OMIT{
\anthony{Lukas, can you put precisely the size here? We want to say something
like $O(c^n |\varphi|)$ for us vs. $O(|\varphi|^n)$ for \cite{LB16,Abdulla14},
where $n$ is the number of splittings we have to do. The second one obviously 
grows faster than the first.}
}

\OMIT{
Our symbolic representation of transitions is more succinct (exponentially in the limit \lukas{hope it is true}) while preserving the worst case complexity of emptiness check and basic automata constructions.
\lukas{is it too much fuss about this?}
}

The succinctness of our AFA representation of string formulae is not for free 
since AFA language emptiness is a PSPACE-complete problem (in contrast to
polynomial-time for NFA). 
%
%
However, 
modern model checking algorithms and heuristics can be harnessed to solve 
the emptiness problem.  
In particular,
we use a linear-time reduction to reachability in Boolean transition systems 
similar to \cite{fang-yu-circuits,arlen}, which can be solved by state of the 
art model checking algorithms, such as IC3 \cite{Bra12}, $k$-induction \cite{DBLP:conf/fmcad/SheeranSS00}, or Craig
interpolation-based methods~\cite{McMillan03}, and tools like nuXmv \cite{nuXmv} or ABC \cite{abc}.

An interesting by-product of our approach is an efficient decision
procedure for the acyclic fragment.  The acyclic logic does not a
priori allow concatenation, but is more liberal in the use of
transducer constraints (which can encode complex relations like
string-length comparisons, and the subsequence relation).  In
addition, such a logic is of interest in the investigation of complex
path-queries for graph databases \cite{BFL13,BLLW12}, which has been
pursued independently of strings for verification.  Our algorithm also
yields an alternative and substantially simpler proof of PSPACE upper
bound of the satisfiability problem of the logic.
\OMIT{
The acyclic logic does not a priori allow concatenation, but is more liberal in the use of transducer constraints.
A limited form other kinds of constraints such as word equations and arithmetic constraints on string lengths can be encoded into them.
%
%
%
(although despite the extended syntax, 
acyclic formulae do not allow flexible enough interaction of concatenation and transducers to be as useful  as straight-line formulae in eg. application vulnerability analysis).
}

We have implemented our AFA-based string solver as the
tool~\strsolver, using the infrastructure provided by the SMT solver
Princess~\cite{princess08}, and applying the nuXmv~\cite{nuXmv} and
ABC~\cite{abc} model checkers to analyse succinct AFAs. \strsolver\ is
a~decision procedure for the discussed fragments of straight-line and
acyclic string formulae, and is able to process SMT-LIB input with
CVC4-style string operations, augmented with operations
\texttt{str.replace}, \texttt{str.replaceall}\footnote{\texttt{str.replaceall}
 is the SMT-LIB syntax for the replace-all
operation. On the other hand, \texttt{str.replace} represents the operation of
replacing the \emph{first} occurrence of the given pattern. In case there is no
such occurrence, the string stays intact.}, and arbitrary transducers
defined using sets of mutually recursive functions. \strsolver\ is
therefore extremely flexible at supporting intricate string
operations, including escape operations such as the ones discussed in
Example~\ref{ex:cacm}.
Experiments with string benchmarks drawn from the literature,
including problems with replace, replace-all, and
general transducers, show that \strsolver\ can solve problems that are
beyond the scope of existing solvers, while it is competitive with
other solvers on problems with a simpler set of operations.

\paragraph{Organisation}
We recall relevant notions from logic and automata theory in Section
\ref{sec:prelim}. In Section~\ref{sec:lang}, we define a general string 
constraint language and mention several important decidable restrictions. In 
Section~\ref{sec:AFA}, we recall the
notion of alternating finite-state automata and define a~succinct variant that
plays a crucial role in our decision procedure. In Section \ref{sec:ac}, we
provide a new algorithm for solving the acyclic fragment of the intersection
of rational relations using AFA. In Section \ref{sec:sl}, we provide our
efficient reduction from the straight-line fragment to the acyclic fragment that
exploits AFA constructions. To simplify the presentation of this reduction,
we first introduce in Section \ref{sec:afa_parameters} a syntactic sugar of
the acyclic fragment called acyclic constraints with synchronisation parameters.
In Section \ref{sec:afaSolving}, we provide our reduction from the AFT emptiness
to reachability in a Boolean transition system. Experimental results are
presented in Section \ref{sec:experiments}. Our tool \strsolver\ can be 
obtained from \url{https://github.com/uuverifiers/sloth/wiki}.
Finally, we conclude in Section \ref{sec:conc}. Missing proofs can be found
in the \shortlong{full version}{appendix}.

\section{Preliminaries}
\label{sec:prelim}

\paragraph{Logic} 
Let $\bool = \{0,1\}$ be the set of Boolean values, and $A$
a set of Boolean variables.
We write $\bofo A$ to denote the set of \emph{Boolean formulae}
 over $A$.
%
In this context, we will sometimes treat subsets $A'$ of $A$ as the
corresponding truth assignments  $\{s \mapsto 1\mid s\in
A'\} \cup \{s\mapsto 0\mid s \in A\setminus A'\}$ and write, for instance, $A'
\models \varphi$ for $\varphi \in\bofo A$ if the assignment satisfies $\varphi$.
An \emph{atom} is a Boolean variable; 
a \emph{literal} is either a atom or its negation.
%
A formula is in \emph{disjunctive normal form} (DNF) if it is a disjunction of
conjunctions of literals, and in \emph{negation normal form} (NNF) if
negation only occurs in front of atoms.
We denote the set of variables
in a formula~$\varphi$ by $\Var\varphi$.
We use $\vect x$ to denote sequences $x_1,\ldots,x_n$ of length $|\vect x| =
n$ of propositional variables, and we write $\varphi(\vect x)$ to denote that
$\vect x$ are the variables of $\varphi$.
If we do not fix the order of the variables, we write $\varphi(X)$ for a formula
with $X$ being its set of variables.
%
For a variable vector $\vect x$, we denote by $\{ \vect x \}$ the set of
variables in the vector.

We say that $\varphi$ is \emph{positive} (\emph{negative})
 on an atom $\alpha\in A$ if  
$\alpha$ appears under an even (odd) number of negations only.
%
A formula that is positive (negative) on all its atoms is called
positive (negative), respectively.
The constant formulae $\true$ and $\false$ are both positive
and negative.
We use $\pobofo S$ and $\nebofo S$ to denote the sets of all positive and
negative Boolean formulae over $S$, respectively.

Given a formula $\varphi$, we write $\polarity{\varphi}$ to denote a
formula obtained by replacing (1) every conjunction by a disjunction
and vice versa and (2) every occurrence of $\true$ by $\false$ and
vice versa. Note that $\polarity{x} = x$, which means that
$\polarity{\varphi}$ is not the same as the negation of $\varphi$.

\paragraph{Strings and languages} 
Fix a finite alphabet $\Sigma$.
Elements in $\Sigma^*$ are interchangeably called words or strings, where the
empty word is denoted by $\epsilon$.  The concatenation of strings $u$, $v$ is denoted
by $u \concat v$, occasionally just by $uv$ to avoid notational clutter. 
We denote by $|w|$ the lenght of a word $w \in \Sigma^*$.
For any word $w=a_1 \ldots a_n$, $n\geq 1$, and any index $1 \leq i \leq n$, we
denote by $\letterof{w}{i}$ the letter $a_i$.
A language is a subset of $\Sigma^*$.  The concatenation of two languages $L,L'$
is the language $L\concat L' = \{w\concat w'\mid w\in L \land w'\in L'\}$, and
the iteration $L^*$ of a language $L$ is the smallest language closed under
$\concat$ and containing $L$ and $\epsilon$.  

\paragraph{Regular languages and rational relations}
A regular language over a finite alphabet $\Sigma$ is a subset of $\Sigma^*$ that can be built by a finite number of applications of the operations of concatenation, iteration, and union 
from the languages $\{\epsilon\}$ and $\{a\}, a\in\Sigma$.
An $n$-ary rational relation $R$ over $\Sigma$ is a subset of $(\Sigma^*)^n$ that can be obtained from a regular language $L$ over the alphabet of $n$-tuples $(\Sigma\cup\{\epsilon\})^n$ as follows.
Include $(w_1,\ldots,w_n)$ in $R$ iff for some
$(a_1^1,\ldots,a_n^1),\ldots,(a_1^k,\ldots,a_n^k)\in L$, $w_i = a_1\circ \cdots
\circ a_k$ for all $1\leq i \leq n$. Here, $\circ$ is a concatenation over the
alphabet $\Sigma$, and $k$ denotes the length of the words $w_i$.
In practice, regular languages and rational relations can be represented
using various flavours of finite-state automata, which are discussed in
detail in Section~\ref{sec:AFA}.

\section{String Constraints}
\label{sec:lang}
We start by recalling a general string constraint language from
\cite{LB16} that supports concatenations, finite-state transducers,
and regular expression matching. 
%
We will subsequently state decidable fragments of the language for
which we design our decision procedure.


\subsection{{\bf String Language}}

We assume a vocabulary
 of countably many {\em string variables}
$x,y,z,\dots$ ranging over $\Sigma^*$.
A \emph{string formula} over $\Sigma$ is a Boolean
combination~$\varphi$ of \emph{word equations}~$x = t$ whose
right-hand side~$t$ might contain the concatenation operator,
\emph{regular constraints}~$\afare(x)$, and \emph{rational
  constraints}~$\aftrel(\vect x)$:
\begin{equation*}
  \varphi ~::=~
  x = t \mid
  \afare(x) \mid
  \aftrel(\vect x) \mid
  \varphi\land\varphi \mid
  \varphi\lor \varphi \mid
  \neg \varphi,
  \qquad
  t ~::=~
  x \mid
  a \mid
  t \circ t~.
\end{equation*}
In the grammar, $x$ ranges over string variables, $\bar x$ over
vectors of string variables, and $a \in \Sigma$ over letters.
$R\subseteq (\Sigma^*)^{n}$ is assumed to be an $n$-ary rational
relation on words of $\Sigma^*$, and $P\subseteq \Sigma^*$ is a
regular language.
We will represent regular languages and rational relations by succinct automata
and transducers denoted as $\aft$ and $\afa$, respectively.
The automata and transducers will be formalized in Section~\ref{sec:AFA}.  
When the transducer $\aft$ or automaton $\afa$ representing a rational relation
$R$ or regular language $P$ is known, we write $\aft(\vect x)$ or
$\afa(\vect x)$ instead of $R(\vect x)$ or $P(\vect x)$ in the formulae,
respectively.

A formula $\varphi$ is interpreted over an \emph{assignment}
$\sass:\Var\varphi\rightarrow \Sigma^*$ of its variables to strings over
$\Sigma^*$.
It \defn{satisfies} $\varphi$, written $\iota\models\varphi$, iff the constraint
$\varphi$ becomes true under the substitution of each variable $x$ by
$\iota(x)$.
We formalise the satisfaction relation for word equations, rational constraints,
and regular constraints, assuming the standard meaning of Boolean
connectives:\begin{enumerate}

\item $\iota$ satisfies the equation $x = t$ if $\iota(x) = \iota(t)$,
  extending $\iota$ to terms by setting $\iota(a) = a$ and $\iota(t_1
  \circ t_2) = \iota(t_1) \circ \iota(t_2)$.

  \item $\iota$ satisfies the rational constraint $\aftrel(x_1,\ldots,x_n)$
    iff
  $(\sass(x_1),\ldots,\sass(x_{n}))$ belongs to $\aftrel$.

  \item $\iota$ satisfies the regular constraint $P(x)$, for $P$ a regular
  language, if and only if $\iota(x) \in P$.

\end{enumerate} 

A satisfying assignment for $\varphi$ is also called a \defn{solution} for 
$\varphi$. If $\varphi$ has a solution, it is 
\defn{satisfiable}.

The unrestricted string logic is undecidable, e.g., one can easily encode
Post Correspondence Problem (PCP) as the problem of checking satisfiability
of the constraint $\aftrel(x,x)$, for some rational transducer $\aftrel$ 
\cite{Morvan00}.
%
We therefore concentrate on practical decidable fragments.

\subsection{Decidable Fragments}
\label{sec:logic}
Our approach to deciding string formulae is based on two major insights. 
The first insight is that alternating automata can be used to efficiently decide positive Boolean combinations of rational constraints. 
This yields an algorithm for deciding (an extension of) the \emph{acyclic 
fragment} of \cite{BFL13}.
The minimalistic definition of acyclic logic restricts rational constraints 
and does not allow word equations 
(in Section~\ref{sec:extensions}
a limited form of equations and arithmetic constraints over lengths 
will be shown to be encodable in the logic).
Our definition of the acyclic logic $\AC$ below generalises that of 
\cite{BFL13} by allowing $k$-ary rational constraints instead of \emph{binary}.

\begin{definition}[Acyclic formulae]
Particularly, we say that a string formula $\varphi$ is \defn{acyclic} if it does not contain word equations, 
rational constraints $\aftrel(x_1,\ldots,x_n)$ only appear positively
and their variables $x_1,\ldots,x_n$ are pairwise distinct,
and
for every sub-formula $\psi \wedge \psi'$ at a positive position of $\varphi$ 
(and also every $\psi \lor \psi'$ at a negative position)
it is the case that $|\Free{\psi} \cap \Free{\psi'}| \leq 1$, i.e., $\psi$ and 
$\psi'$ have \emph{at most one} variable in common.
We denote by $\AC$ the set of all acyclic formulae.
\end{definition}
%

The second main insight we build on is that alternation allows a very efficient encoding of concatenation into rational constraints and automata (though only equisatisfiable, not equivalent).
Efficient reasoning about concatenation combined with rational relations is the main selling point of our work from the practical perspective---this is what is most needed and was so far missing in applications like security analysis of web-applications.
We follow the approach from \cite{LB16} which defines so called straight-line conjunctions. 
Straight-line conjunctions essentially correspond to sequences of program assignments in the single static assignment form, possibly interleaved with assertions of regular properties. 
An equation $x = y_1\concat\cdots\concat y_n$ is understood as an assignment to a program variable $x$. 
A rational constraint $\aftrel(x,y)$ may be interpreted as an assignment to $x$ as well,
in which case we write it as $x = \aftrel(y)$
(though despite the notation, $\aftrel$ is not required to represent a function, it can still mean any rational relation). 
%
%
\begin{definition}[Straight-line conjunction]
A conjunction of string constraints is then defined to be \emph{straight-line} if it can be written as $\psi\land\bigwedge_{i=1}^m x_i = P_i$
where $\psi$ is a conjunction of regular and negated regular constraints and each $P_i$ is either of the form $y_1\concat\cdots\concat y_n$, or $R(y)$ and, importantly, 
$P_i$ cannot contain variables $x_{i},\ldots, x_m$. 
We denote by $\SL$ the set of all straight-line conjunctions.
\end{definition}

\begin{example}
\label{ex:cacmSL}
The program snippet in Example~\ref{ex:cacm} would be expressed as $x = \aftrel_1(\mathsf{name})\land y = \aftrel_2(x) \land z = w_1 \concat y \concat w_2 \concat x \concat w_3\land  u = \aftrel_3(z)$. 
The transducers $\aftrel_i$ correspond to the string operations at the respective lines: 
$\aftrel_1$ is the $\mathsf{htmlEscape}$, 
$\aftrel_2$ is the $\mathsf{escapeString}$, and 
$\aftrel_3$ is the implicit transduction within $\mathsf{innerHTML}$.  
%
Line 3 is translated into a conjunction of the concatenation and the third rational constraint encoding the implicit string operation at the assignment to $\mathsf{innerHTML}$. 
In the concatenation, $w_1,w_2,w_3$ are words that correspond to the three constant strings concatenated with $x$ and $y$ on line 3. 
To test vulnerability, a regular constraint $\afa(u)$ encoding the 
    \mbox{pattern $\mathsf{e1}$ is added as a conjunct.}
\end{example}
%
%
The fragment of straight-line conjunctions can be straightforwardly extended to disjunctive formulae. We say that a string formula is straight-line if every clause in its DNF
is straight-line. 
%
A decision procedure for straight-line conjunctions immediately extends to
straight-line formulae: instantiate the DPLL(T) framework \cite{tinelli_dpll_2004} with a solver for straight-line conjunctions.

The straight-line and acyclic fragments are clearly syntactically incomparable: $\AC$ does not have equations, $\SL$ restricts more strictly combinations of rational relations and allows only binary ones.  
Regarding expressive power, $\SL$ can express properties which $\AC$ cannot: 
the straight-line constraint $x = yy$ cannot be expressed by any acyclic formula.
On the other hand, whether or not $\AC$ formulae can be expressed in $\SL$ is not clear. 
Every $\AC$ formula can be expressed by a~single $n$-ary acyclic rational constraint (c.f. Section~\ref{sec:ac}), 
hence acyclic formulae and acyclic rational constraints are of the same power. 
It is not clear however whether straight-line formulae, which can use only binary rational constraints, can  express arbitrary $n$-ary acyclic rational constraint. 

\section{Succinct Alternating Automata and Transducers} \label{sec:AFA}

We introduce a~succinct form of alternating automata and transducers that
operate over \emph{bit vectors}, i.e., functions $\bv: \bvar \rightarrow \bool$
where $\bvar$ is a finite, totally ordered set of bit variables.
This is a variant of the recent automata model in 
\cite{dantoni:alternating} that is tailored to our problem.
%
%
Bit vectors can of course be described by strings over $\bool$,
conjunctions of literals over $\bvar$, or sets of those elements $v \in \bvar$
such that $\bv(v) = 1$.
In what follows, we will use all of these representations interchangeably.
Referring to the last mentioned possibility, we denote the set of all bit
vectors over $\bvar$ by $\BV$.

An obvious advantage of this approach is that encoding symbols of large
alphabets, such as UTF, by bit vectors allows one to succinctly represent sets
of such symbols using Boolean formulae. 
In particular, symbols of an alphabet of size $2^k$ can be encoded by bit
vectors of size $k$ (or, alternatively, as Boolean formulae over $k$ Boolean
variables).
We use this fact when encoding transitions of our alternating automata.

\begin{example}\label{ex:alphabet} To illustrate the encoding, assume
 the alphabet $\Sigma = \{ a, b, c, d \}$ consisting of symbols
$a$, $b$, $c$, and $d$. We can deal with this alphabet by using the set $\bvar =
\{ v_0, v_1 \}$ and representing, e.g., $a$ as $\neg v_1 \wedge \neg v_0$, $b$
as $\neg v_1 \wedge v_0$, $c$ as $v_1 \wedge \neg v_0$, and $d$ as $v_1 \wedge
v_0$. This is, $a$, $b$, $c$, and $d$ are encoded as the bit vectors $00$, $01$,
$10$, and $11$ (for the ordering $v_0 < v_1$), or the sets $\emptyset$,
$\{v_0\}$, $\{v_1\}$, $\{v_0,v_1\}$, respectively. The set of symbols $\{ c, d
\}$ can then be encoded simply by the formula $v_1$.\qed\end{example}

\subsection{Succinct Alternating Finite Automata}

A \emph{succinct alternating finite automaton (AFA)} over Boolean variables
$\bvar$ is a tuple $\afa = (\bvar,\states,\Delta,\init,\final)$ where $\states$
is a finite set of \emph{states}, the \emph{transition function} $\Delta:
\states \to \bofo{\bvar\cup\states}$ assigns to every state a~Boolean formula
over Boolean variables and states that is positive on states, $\init \in
\pobofo{\states}$ is a positive \emph{initial formula}, and $\final \in
\nebofo{\states}$ is a negative \emph{final formula}. 
Let $w = \symb_1\ \ldots \symb_m$, $m \geq 0$, be a word where each $b_i$, $1
\leq i \leq m$, is a bit vector encoding the $i$-th letter of $w$.
A \emph{run} of the AFA $\afa$ over $w$ is a~sequence $\run =
\conf_0\symb_1\conf_1\ldots\symb_{m}\conf_m$ where $b_i \in \BV$ for every $1
\leq i \leq m$, $\conf_i \subseteq \states$ for every $0 \leq i \leq m$, and
$\symb_i \cup \conf_i \models \bigwedge_{q\in \conf_{i-1}}\Delta(q)$ for every
$1 \leq i \leq m$. 
The run is \emph{accepting} if $\conf_0 \models I$ and $\conf_m\models F$, in
which case the word is accepted. The \emph{language} of $\afa$ is the set
$L(\afa)$ of accepted words.

Notice that instead of the more usual definition of $\Delta$, which would assign
a positive Boolean formula over $\states$ to every pair from $\states\times\BV$
or to a pair $\states\times\bofo{\bvar}$ as in \cite{dantoni:alternating}, we
let $\Delta$ assign to states formulae that talk about both target states and
Boolean input variables. This is closer to the encoding of the transition
function as used in MONA \cite{monasecrets}. It allows for additional
succinctness and also for a more natural translation of the language emptiness
problem into a model checking problem (cf. Section~\ref{sec:afaSolving}).\footnote{\cite{dantoni:alternating} also mentions an implementation of symbolic AFAs that uses MONA-like BDDs and is technically close to our AFAs.}
Moreover, compared with the usual AFA definition, we do not have just a single
initial state and a single set of accepting states, but we use initial and final
formulae. As will become clear in Section~\ref{sec:ac}, this approach allows us
to easily translate the considered formulae into AFAs in an inductive way.

Note that standard \emph{nondeterministic finite automata} (NFAs), working over
bit vectors, can be obtained as a special case of our AFAs as follows. An AFA
$\afa = (\bvar,\states,\Delta,\init,\final)$ is an NFA iff (1) $\init$ is of the
form $\bigvee_{q \in Q'} q$ for some $Q' \subseteq Q$, (2) $\final$ is of the
form $\bigwedge_{q \in Q''} \neg q$ for some $Q'' \subseteq Q$, and (3) for
every $q \in Q$, $\Delta(q)$ is of the form $\bigvee_{1 \leq i \leq m}
\varphi_i(\bvar) \wedge q_i$ where $m\geq 0$ and, for all $1 \leq i \leq m$,
$\varphi_i(\bvar)$ is a~formula over the input bit variables and $q_i \in Q$.

\begin{example}\label{ex:AFA} To illustrate our notion of AFAs, we give an
example of an AFA $\afa$ over the alphabet $\Sigma = \{ a,b,c,d \}$ from
Example~\ref{ex:alphabet} that accepts the language $\{ w \in \Sigma^* \mid |w|
\nicemod 35 = 0 ~\wedge$
%
%
$\forall i \exists j: (1 \leq i \leq |w| \wedge w[i] \in \{a,b\}) \rightarrow (i
< j \leq |w| \wedge w[j] \in \{c,d\})\}$, i.e., the length of the words is 
a~multiple of $35$, and every letter $a$ or $b$ is eventually followed by a letter
$c$ or $d$. In particular, we let $\afa = ( \{v_0, v_1\},
\{q_0,\ldots,q_4,p_0,\ldots,p_6,r_1,r_2\} \}, \Delta, \init, \final)$ where
$\init = q_0 \wedge p_0$, $\final = \neg q_1 \wedge \ldots \wedge \neg q_4
\wedge \neg p_1 \wedge \ldots \wedge \neg p_6 \wedge \neg r_1$ (i.e., the
accepting states are $q_0$, $p_0$, and $r_2$), and $\Delta$ is defined as
follows:\begin{itemize}

  \item $\forall 0 \leq i < 5: \Delta(q_i) = (\neg v_1 \wedge q_{(i+1) \nicemod
  5} \wedge r_1) \vee (v_1 \wedge q_{(i+1) \nicemod 5})$,

  \item $\forall 0 \leq i < 7: \Delta(p_i) = p_{(i+1) \nicemod 7}$,

  \item $\Delta(r_1) = (v_1 \wedge r_2) \vee (\neg v_1 \wedge r_1)$ and
  $\Delta(r_2) = r_2$.

\end{itemize} Intuitively, the $q$ states check divisibility by $5$. Moreover,
whenever, they encounter an $a$ or $b$ symbol (encoded succinctly as checking
$\neg v_1$ in the AFA), they spawn a run through the $r$ states, which checks
that eventually a $c$ or $d$ symbol appears. The $p$ states then check
divisibility by $7$. The desired language is accepted due to the requirement
that all these runs must be synchronized. Note that encoding the language using
an NFA would require quadratically more states since an explicit product of all
the branches would have to be done.\qed\end{example}


The additional succinctness of AFA does not
influence the computational complexity of the emptiness check compared to the standard variant of alternating automata.

\begin{lemma}
The problem of language emptiness of AFA is PSPACE-complete.
\end{lemma}

The lemma is witnessed by a linear-space transformation of the
problem of emptiness of an AFA language to the PSPACE-complete problem of
reachability in a Boolean transition system. This transformation is shown in
Section~\ref{sec:afaSolving}.

\subsection{Boolean Operations on AFAs}\label{sec:AFA-ops}

From the standard Boolean operations over AFAs, we will mainly need conjunction
and disjunction in this paper. These operations can be implemented in linear
space and time in a~way analogous to \cite{dantoni:alternating}, slightly
adapted for our notion of initial/final formulae, as follows. 
Given two AFAs $\afa = (\bvar,\states,\Delta,\init,\final)$ and $\afa' =
(\bvar,\states',\Delta',\init',\final')$ with $\states\cap\states'=\emptyset$,
the automaton accepting the union of their languages can be constructed as
$\afa\cup\afa' =
(\bvar,\states\cup\states',\transitions\cup\transitions',\init\lor\init',\final\land\final')$,
and the automaton accepting the intersection of their languages can be
constructed as $\afa\cap\afa' =
(\bvar,\states\cup\states',\transitions\cup\transitions',\init\land\init',\final\land\final')$.
Seeing correctness of the construction of $\afa\cap\afa'$ is immediate. Indeed,
the initial condition enforces that the two AFAs run in parallel, disjointness
of their state-spaces prevents them from influencing one another, and the final
condition defines their parallel runs as accepting iff both of the runs accept.
To see correctness of the construction of $\afa\cup\afa'$, it is enough to
consider that one of the automata can be started with the empty set of states
(corresponding to the formula $\bigwedge_{q \in Q} \neg q$ for $\afa$ and
likewise for $\afa'$). This is possible since only one of the initial formulae
$\init$ and $\init'$ needs to be satisfied. The automaton that was started with
the empty set of states will stay with the empty set of states throughout the
entire run and thus trivially satisfy the (negative) final formula. 

\begin{example} Note that the AFA in Example \ref{ex:AFA} can be viewed as
obtained by conjunction of two AFAs: one consisting of the $q$ and $r$ states
and the second of the $p$ states. \qed\end{example}

To complement an AFA $\afa = (\bvar,Q,\Delta,\init,\final)$,
we first transform the automaton into a form corresponding to the symbolic AFA of \cite{dantoni:alternating} and then use their complementation procedure. 
More precisely, the transformation to the symbolic AFA form requires two steps:
\begin{itemize}
  \item
  The first step simplifies the final condition.
  The final formula $\final$ is converted into DNF, yielding a
  formula $\final_1 \vee \ldots \vee \final_k$, $k \geq 1$, where each
  $\final_i$, $1 \leq i \leq k$, is a conjunction of negative literals over $Q$.
  The AFA
  $\afa$ is then transformed into a union of AFAs $\afa_i =
  (\bvar,Q,\Delta,\init,\final_i)$, $1 \leq i \leq k$, where each 
  $\afa_i$ is a copy of $\afa$ except that it uses one of the disjuncts 
  $\final_i$ of the DNF form of
  the original final formula $\final$ as its final formula.
%
%
Each resulting AFAs hence have a purely conjunctive final condition that corresponds a set of final states of \cite{dantoni:alternating} 
(a~set of final states $F\subseteq Q$ would correspond to the final formula $\bigwedge_{q \in Q \setminus F} \neg q$).

  \item 
  The second step simplifies the structure of the transitions.
  For every $q \in Q$, the transition formula $\Delta(q)$ is
  transformed into a disjunction of formulae of the form $(\varphi_1(\bvar)
  \wedge \psi_1(Q)) \vee \ldots \vee (\varphi_m(\bvar) \wedge \psi_m(Q))$ where
  the $\varphi_i(\bvar)$ formulae, called \emph{input formulae} below, speak
  about input bit variables only, while the $\psi_i(Q)$ formulae, called
  \emph{target formulae} below, speak exclusively about the target states, for
  $1 \leq i \leq m$. For this transformation, a slight modification of
  transforming a formula into DNF can be used. 
\end{itemize}

The complementation procedure of \cite{dantoni:alternating} then proceeds in two steps: the \emph{normalisation} and the complementation itself. We sketch them below:
\begin{itemize}
  
  \item For every $q \in
  Q$, normalisation transforms the transition formula $\Delta(q) = (\varphi_1(\bvar) \wedge \psi_1(Q))
  \vee \ldots \vee (\varphi_m(\bvar) \wedge \psi_m(Q))$
  so that every two distinct input formulae $\varphi(\bvar)$ and
  $\varphi'(\bvar)$ of the resulting formula describe disjoint sets of bit
  vectors, i.e., $\neg (\varphi(\bvar) \wedge \varphi'(\bvar))$ holds. To
  achieve this (without trying to optimize the algorithm as in
  \cite{dantoni:alternating}), one can consider generating all Boolean
  combinations of the original $\varphi(\bvar)$ formulae, conjoining each of
  them with the disjunction of those state formulae whose input formulae are
  taken positively in the given case. More precisely, one can take $\bigvee_{I
  \subseteq \{1,..., m\}} (\bigwedge_{i \in I} \varphi_i)) \wedge (\bigwedge_{i
  \in \{1, \ldots, m \} \setminus I} \neg \varphi_i)) \wedge \bigvee_{i \in I}
  \psi_i$. 
  
  \item Finally, to complement the AFAs normalized in the above way, one
  proceeds as follows: (1)~The initial formula $\init$ is replaced by
  $\polarity{\init}$. (2)~For every $q \in Q$ and every disjunct $\varphi(\bvar)
  \wedge \psi(Q)$ of the transition formula $\Delta(q)$, the target formula
  $\psi(Q)$ is replaced by $\polarity{\psi}(Q)$. (3)~The final formula of the
  form $\bigwedge_{q \in Q'} \neg q$, $Q' \subseteq Q$, is transformed to the
  formula $\bigwedge_{q \in Q \setminus Q'} \neg q$, and $\mathtt{false}$ is
  swapped for $\mathtt{true}$ and vice versa.

\end{itemize}

Clearly, the complementation contains three sources of exponential blow-up: 
(1)
the simplification of the final condition, 
(2)
the simplification of transitions
and
(3)
the normalization of transitions.
Note, however, that, in this paper, we will apply
complementation exclusively on AFAs obtained by Boolean operations from NFAs
derived from regular expressions. Such AFAs already have the simple final conditions, and
so the first source of exponential blow-up does not apply. The second and the third source of
exponential complexity can manifest themselves but note that it does not show up in the number of
states. Finally, note that if we used AFAs with explicit
alphabets, the second and the third problem would disappear (but then the AFAs would usually be
bigger anyway).

\subsection{Succinct Alternating Finite Transducers} 

In our alternating finite transducers, we will need to use \emph{epsilon
symbols} representing the empty word. Moreover, as we will explain later, in
order to avoid some undesirable synchronization when composing the transducers,
we will need more such symbols---differing just syntactically. Technically, we
will encode the epsilon symbols using a set of epsilon bit variables $\eps$,
containing one new bit variable for each epsilon symbol. We will draw the
epsilon bit variables from a countably infinite set $\epsuni$.
 We will also
assume that when one of these bits is set, other bits
 are not important.

Let $\bvare$ be a finite, totally ordered set of bit variables, which we can
split to the set of input bit variables $\bvar(\bvare) = \bvare \setminus
\epsuni$ and the set of epsilon bit variables $\eps(\bvare) = \bvare \cap
\epsuni$.
Given a word $w = b_1 \ldots b_m \in \power{\bvare}^*$, $m \geq 0$, we denote by
$\compact{w}$ the word that arises from $w$ by erasing all those $b_i$, $1 \leq
i \leq m$, in which some epsilon bit variable is set, i.e., $b_i \cap \epsuni
\neq \emptyset$.
Further, let $k\geq 1$, and let $\bvarek = \bvare \times [k]$, assuming it to be
ordered in the lexicographic way.
The indexing of the bit variables will be used to express the track on which
they are read.
Finally, given a word $w  = b_1 \ldots b_m \in \power{\bvarek}^*$, $m \geq 0$,
we denote by $\select{w}{i}$, $1 \leq i \leq k$, the word $b'_1 \ldots b'_m \in
\power{\bvare}^*$ that arises from $w$ by keeping the contents of the $i$-th
track (without the index~$i$) only, i.e., $b'_j \times \{ i \} = b_j \cap
(\bvare \times \{ i \})$ for $1 \leq j \leq m$.

A $k$-track \emph{succinct alternating finite transducer} (AFT) over $\bvare$ is
syntactically an alternating automaton $\aft =
(\bvarek,\states,\Delta,\init,\final)$, $k \geq 1$.
Let $\bvar = \bvar(\bvare)$.
The relation $\rel(\aft) \subseteq (\power{\bvar}^*)^k$ \emph{recognised} by
$\aft$ contains a $k$-tuple of words $(x_1, \ldots, x_k)$ over $\BV$ iff there
is a word $w \in L(\aft)$ such that $x_i =~ \compact{\select{w}{i}}$ for each $1
\leq i \leq k$.

Below, we will sometimes say that the word  $w$ encodes the $k$-tuple of words
$(x_1, \ldots, x_k)$.
Moreover, for simplicity, instead of saying that $\aft$ has a run over $w$ that
encodes $(x_1, \ldots, x_k)$, we will sometimes directly say that $\aft$ has a
run over $(x_1, \ldots, x_k)$ or that $\aft$ accepts $(x_1, \ldots, x_k)$.

Finally, note that classical \emph{nondeterministic finite transducers} (NFTs)
are a special case of our AFTs that can be defined by a similar restriction as
the one used when restricting AFAs to NFAs. 
In particular, the first track (with letters indexed with $1$) can be seen as
the input track, and the second track (with letters indexed with $2$) can be
seen as the output track.
AFTs as well as NFTs recognize the class of \emph{rational relations}
\cite{BFL13,Berstel,Sakarovitch}.

\begin{example}\label{ex:AFT} We now give a simple example of an AFT that
implements escaping of every apostrophe by a backlash in the UTF-8 encoding.
%
%
%
Intuitively, the AFT will transform an input string \verb|x'xx| to the
string \verb|x\'xx|, i.e., the relation it represents will contain the couple
$($\verb|x'xx|$,$\verb|x\'xx|$)$.
All the symbols should, however, be encoded in UTF-8.
In this encoding, the apostrophe has the binary code $00100111$, and the
backlash has the code $00101010$. We will work with the set of bit variables
$\bvar_8 = \{ v_0, \ldots, v_7 \}$ and a single epsilon bit variable $e$. We
will superscript the bit variables by the track on which they are read (hence,
e.g., $v_1^2$ is the same as $(v_1,2)$, i.e., $v_1$ is read on the second
track). Let $\mathtt{ap}^i = v_0^i \wedge v_1^i \wedge v_2^i \wedge \neg v_3^i
\wedge \neg v_4^i \wedge v_5^i \wedge \neg v_6^i \wedge \neg v_7^i \wedge \neg
e^i$ represent an apostrophe read on the $i$-th track. Next, let $\mathtt{bc}^i
= \neg v_0^i \wedge v_1^i \wedge \neg v_2^i \wedge v_3^i \wedge \neg v_4^i
\wedge v_5^i \wedge \neg v_6^i \wedge \neg v_7^i \wedge \neg e^i$ represent 
a~backlash read on the $i$-th track. Finally, let $\mathtt{eq}^{i,j} = e^i
\leftrightarrow e^j \wedge \bigwedge_{0 \leq k < 8} v^i_k \leftrightarrow v^j_k$
denote that the same symbol is read on the $i$-th and $j$-th track. The AFT that
implements the described escaping can be constructed as follows: $\aft =
(\bvarxx{(\bvar_8 \cup \{ e \})}{2},\{ q_0, q_1 \},\Delta,q_0,\neg q_1)$ where
the transition formulae are defined by
   $\Delta(q_0) = (\neg \mathtt{ap}^1 \wedge \mathtt{eq}^{1,2} \wedge q_0) \vee
   (\mathtt{ap}^1 \wedge \mathtt{bc}^2 \wedge q_1$) and
   $\Delta(q_1) = e^1 \wedge \mathtt{ap}^2 \wedge q_0$.  \qed \end{example}


\section{Deciding Acyclic Formulae} \label{sec:ac}


Our decision procedure for AC formulae is based on translating them into AFTs.
For simplicity, we assume that the formula is negation free 
(after transforming to NNF, negation at regular constraints can be eliminated by AFA complementation).
Notice that with no negations, the restriction $\AC$ puts on disjunctions never applies. 
We also assume that the formula contains rational constraints only 
(regular constraint can be understood as unary rational constraints).

Our algorithm then transforms a formula $\varphi(\vect x)$ into a rational constraint
$\aft_\varphi(\vect x)$ inductively on the structure of $\varphi$.
As the base case, we get rational constraints $\aft(\vect x)$, which are already
represented as AFTs, and regular constraints $\afa(x)$, already represented by
AFAs.
Boolean operations over regular constraints can be treated using the
corresponding Boolean operations over AFAs described in
Section~\ref{sec:AFA-ops}.
The resulting AFAs can then be viewed as rational constraints with one variable
(and hence as a~single-track AFT).

Once constraints $\aft_\varphi(\vect x)$ and $\aft_\psi(\vect y)$ are available,
the induction step translates formulae $\aft_\varphi(\vect x) \wedge
\aft_\psi(\vect y)$ and $\aft_\varphi(\vect x) \vee \aft_\psi(\vect y)$ to
constraints $\aft_{\varphi\land\psi}(\vect z)$ and $\aft_{\varphi\lor\psi}(\vect
z)$, respectively.
To be able to describe this step in detail, let $\aft_\varphi = (\bvarxx{(\bvar
\cup \eps_\varphi)}{|\vect
x|},\controls_\varphi,\transrel_\varphi,\init_\varphi,\final_\varphi)$ and
$\aft_\psi = (\bvarxx{(\bvar \cup \eps_\psi)}{|\vect
y|},\controls_\psi,\transrel_\psi,\init_\psi,\final_\psi)$ such that w.l.o.g.
$\states_\varphi\cap\states_\psi=\emptyset$ and $\eps_\varphi \cap \eps_\psi =
\emptyset$. 

\paragraph{Translation of conjunctions to AFTs.}

The construction of $\aft_{\varphi\land\psi}$ has three steps:\begin{enumerate}

  \item \textbf{Alignment of tracks} that ensures that distinct variables are
  assigned different tracks and that the transducers agree on the track used for
  the shared variable.

  \item \textbf{Saturation by $\epsilon$-self loops} allowing the AFTs to
  synchronize whenever one of them makes an $\epsilon$ move on the shared track.

  \item \textbf{Conjunction} on the resulting AFTs viewing them as AFAs.
  
\end{enumerate}

\paragraph{Alignment of tracks.}

Given constraints $\aft_\varphi(\vect x)$ and $\aft_\psi(\vect y)$, the goal of
the alignment of tracks is to assign distinct tracks to distinct variables of
$\vect x$ and $\vect y$, and to assign the same track in both of the transducers
to the shared variable---if there is one (recall that, by acyclicity, $\vect x$
and $\vect y$ do not contain repeating variables and share at most one common
variable).
This is implemented by choosing a~vector $\vect z$ that consists of exactly one
occurrence of every variable from $\vect x$ and $\vect y$, i.e., $\{\vect z\} =
\{\vect x\} \cup \{\vect y\}$, and by subsequently re-indexing the bit vector
variables in the transition relations.
Particularly, in $\transitions_\varphi$, every indexed bit vector variable $v^i$
(including epsilon bit variables) is replaced by $v^j$ with $j$ being the
position of $x_i$ in $\vect z$, and analogously in  $\transitions_\psi$, every
indexed bit variable $v^i$ is replaced by $v^j$ with $j$ being the position of
$y_i$ in $\vect z$.
Both AFTs are then considered to have $|\vect z|$ tracks.

\paragraph{Saturation by $\epsilon$-self loops.}

This step is needed if $\vect x$ and $\vect y$ share a variable, i.e., $\{\vect
x\} \cap \{\vect y\} \neq \emptyset$.
The two input transducers then have to synchronise on reading its symbols.
However, it may happen that, at some point, one of them will want to read from
the non-shared tracks exclusively, performing an $\epsilon$ transition on the
shared track.
Since reading of the non-shared tracks can be ignored by the other transducer,
it should be allowed to perform an $\epsilon$ move on all of its tracks.
However, that needs not be allowed by its transition function.
To compensate for this, we will saturate the transition function by
$\epsilon$-self loops performed on all tracks.
Unfortunately, there is one additional problem with this step: If the added
$\epsilon$ transitions were based on the same epsilon bit variables as those
already used in the given AFT, they could enable some additional synchronization
\emph{within} the given AFT, thus allowing it to accept some more tuples of
words.
We give an example of this problem below (Example~\ref{ex:AFT-intersection}).
To resolve the problem, we assume that the two AFTs being conjuncted use
different epsilon bit variables (more of such variables can be used due the AFTs
can be a result of several previous conjunctions).
Formally, for any choice $\sigma,\sigma' \in\{\varphi,\psi\}$ such that $\sigma
\neq \sigma'$, and for every state $q\in\states_\sigma$, the transition formula
$\transrel_\sigma(q)$ is replaced by $\transrel_\sigma(q)\lor (q \land
\bigvee_{e \in E_{\sigma'}} \bigwedge_{i \in [|\vect z|]} e^i)$.

\paragraph{Conjunction of AFTs viewed as AFAs.}

In the last step, the input AFTs with aligned tracks and saturated by
$\epsilon$-self loops are conjoined using the automata intersection construction
from Section~\ref{sec:AFA-ops}.

\begin{lemma} Let $\aft'_\varphi$ and $\aft'_\psi$ be the AFTs obtained from the
input AFTs $\aft_\varphi$ and $\aft_\psi$ by track alignment and
$\epsilon$-self-loop saturation, and let $\aft_{\varphi\land\psi} =
\aft_\varphi'\cap\aft_\psi'$. 
Then, $\aft_{\varphi\land\psi}(\vect z)$ is equivalent to 
$\aft_\varphi(\vect x) \wedge \aft_\psi(\vect y)$. \end{lemma}

To see that the lemma holds, note that both $\aft'_\varphi$ and $\aft'_\psi$ have
the same number of tracks---namely, $|\vect z|$. This number can be bigger than
the original number of tracks ($|\vect x|$ or $|\vect y|$, resp.), but the AFTs
still represent the same relations over the original tracks (the added tracks
are unconstrained). The $\epsilon$-self loop saturation does not alter the
represented relations either as the added transitions represent empty words
across all tracks only, and, moreover, they cannot synchronize with the original
transitions, unblocking some originally blocked runs. Finally, due to the
saturation, the two AFTs cannot block each other by an epsilon move on the
shared track available in one~of~them~only.\footnote{Note that the same approach
cannot be used for AFTs sharing more than one track. Indeed, by intersecting
two general rational relations, one needs not obtain a rational relation.}

\begin{example}\label{ex:AFT-intersection} We now provide an example
illustrating the conjunction of AFTs, including the need to saturate the AFTs by
$\epsilon$-self loops with different $\epsilon$ symbols. We will assume working
with the input alphabet $\Sigma = \{a,b\}$ encoded using a single input bit
variable $v_0$: let $a$ correspond to $\neg v_0$ and $b$ to $v_0$. Moreover, we
will use two epsilon bit variables, namely, $e_1$ and $e_2$.
%
%
We consider the following two simple AFTs, each with two tracks:\begin{itemize}

  \item $\aft_1 = (\bvarxx{\{v_0,e_1\}}{2},\{q_0,q_1,q_2\},\Delta_1,q_0,\neg q_0
  \wedge \neg q_2)$ with $\Delta_1(q_0) = (a^1 \wedge b^2 \wedge q_1) \vee (a^1
  \wedge a^2 \wedge q_1 \wedge q_2)$, $\Delta_1(q_1) = \false$, and
  $\Delta_1(q_2) = e_1^1 \wedge q_1$. Note that $\rel(\aft_1) = \{ (a,b) \}$
  since the run that starts with $a^1 \wedge a^2$ gets stuck in one of its
  branches, namely the one that goes to $q_2$. This is because we require
  branches of a single run of an AFT to synchronize even on epsilon bit
  variables, and the transition from $q_2$ cannot synchronize with any move from
  $q_1$. 

  \item $\aft_2 = (\bvarxx{\{v_0,e_2\}}{2},\{p_0,p_1,p_2\},\Delta_2,p_0,\neg p_0
  \wedge \neg p_1)$ such that $\Delta_2(p_0) = (a^1 \wedge b^2 \wedge p_1)$,
  $\Delta_2(p_1) = e_2^1 \wedge b^2 \wedge p_2$, and $\Delta_2(p_2) = \false$.
  Clearly, $\rel(\aft_2) = \{ (a,bb) \}$.

\end{itemize} Let $Q_i$, $\init_i$, $\final_i$ denote the set of states, initial
constraint, and final constraint of $\aft_i$, $i \in \{1,2\}$, respectively.
Assume that we want to construct an AFT for the constraint $\aft_1(x,y) \wedge
\aft_2(x,z)$. This constraint represents the ternary relation $\{ (a,b,bb) \}$.
It can be seen that if we apply the above described construction for
intersection of AFTs to $\aft'_1$ and $\aft'_2$, where $\aft'_1 = \aft_1$ and
$\aft'_2$ is the same as $\aft_2$ up to all symbols from track to 2 are moved to
track 3, we will get an AFT $\aft = (\bvarxx{\{v_0,e_1,e_2\}}{3}, Q_1 \cup Q_2,
\Delta, \init_1 \wedge \init_2, \final_1 \wedge \final_2)$ representing exactly
this relation. We will not list here the entire $\Delta$ but let us note the
below:\begin{itemize}

  \item $\Delta$ will contain the following transition obtained by
  $\epsilon$-self-loop saturation of $\aft_1$: $\Delta(q_1) = (e_2^1 \wedge
  e_2^2 \wedge q_1)$. This will allow $\aft$ to synchronize its run through
  $q_1$ with its run from $p_1$ to $p_2$. Without the saturation, this would not
  be possible, and $\rel(\aft)$ would be empty.

  \item On the other hand, if a single epsilon bit variable $e$ was used in both
  AFTs as well as in their saturation, the saturated $\Delta_1$ would include
  the transition $\Delta_1(q_1) = (e^1 \wedge e^2 \wedge q_1)$. This transition
  could synchronize with the transition $\Delta_1(q_2) = e^1 \wedge q_1$, and
  the relation represented by the saturated $\aft_1$ would grow to $\rel(\aft_1)
  = \{ (a,b), (a,a) \}$. The result of the intersection would then (wrongly)
  represent the relation $\{ (a,b,bb), (a,a,bb) \}$.\qed

\end{itemize}
\end{example}

\paragraph{Translation of disjunctions to AFTs.}

The construction of an AFT for a disjunction of formulae is slightly simpler.
The alignment of variables is immediately followed by an application of the AFA
disjunction construction.
That is, the AFT $\aft_{\varphi\lor\psi}$ is constructed simply as
$\aft_\varphi'\cup\aft_\psi'$ from the constraints $\aft_\varphi'(\vect z)$ and
$\aft_\psi'(\vect z)$ produced by the alignment of the vectors of variables
$\vect x$ and $\vect y$ in $\aft_\varphi(\vect x)$ and $\aft_\psi(\vect y)$.
The construction of $\aft_{\varphi}'$ and $\aft_\psi'$ does not require the
saturation by $\epsilon$-self loops because the two transducers do not need to
synchronise on reading shared variables.
The vectors $\vect x$ and $\vect y$ are allowed to share any number of
variables.

\begin{theorem}\label{th:afa_cons} Every acyclic formula $\varphi(\vect x)$ can
be transformed into an equisatisfiable rational constraint $\aft(\vect x)$
represented by an AFT $\aft$. The transformation can be done in polynomial time
unless $\varphi$ contains a negated regular constraint represented by a
non-normalized succinct NFA.\end{theorem}

\begin{corollary}\label{cor:pspaceAC} Checking satisfiability of acyclic
formulae is in PSPACE unless the formulae contain a~negated regular constraint
represented by a non-normalized succinct NFA.\end{corollary}

PSPACE membership of satisfiability of acyclic formulae with binary rational constraints (without negations of
regular constraints and without considering succinct alphabet encoding) is
proven already in \cite{BFL13}. 
Apart from extending the result to $k$-ary rational constraints, we obtain a
simpler proof as a corollary of Theorem \ref{th:afa_cons}, avoiding a need to use
the highly intricate polynomial-space procedure based on the Savitch\'{}s trick
used in \cite{BFL13}.
Not considering the problem of negating
regular constraints, our PSPACE algorithm would first construct a~linear-size
AFT for the input $\varphi$. We can then use the fact that the standard PSPACE
algorithm for checking emptiness of AFAs/AFTs easily generalises to succinct
AFAs/AFTs. This is proved by our linear-space reduction of emptiness of the
language of succinct AFAs to reachability in Boolean transition systems,
presented in Section~\ref{sec:afaSolving}. Reachability in Boolean transition
systems is known to be PSPACE-complete.


\subsection{Decidable Extensions of AC}
\label{sec:extensions}
The relatively liberal condition that $\AC$ puts on rational constraints allow us to easily extend $\AC$ with other features, without having to change the decision procedure. Namely, we can add Presburger constraints about word length, as well as word equations, as long as overall acyclicity of a formula is preserved.
Length constraints can be added in the general form
$\varphi_{\text{Pres}}(|x_1|, \ldots, |x_k|)$, where
$\varphi_{\text{Pres}}$ is a Presburger formula.
\begin{definition}[Extended acyclic formulae]
  A string formula $\varphi$ augmented with length
  constraints $\varphi_{\text{Pres}}(|x_1|, \ldots, |x_k|)$ is
  \defn{extended acyclic} if every word equation or rational
  constraint contains each variable at most once, rational constraints
  $\aftrel(x_1,\ldots,x_n)$ only appear at positive positions, and for
  every sub-formula $\psi \wedge \psi'$ at a positive position of
  $\varphi$ (and also every $\psi \lor \psi'$ at a negative position)
  it is the case that $|\Free{\psi} \cap \Free{\psi'}| \leq 1$, i.e.,
  $\psi$ and $\psi'$ have \emph{at most one} variable in common.
\end{definition}

Any extended $\AC$ formula $\varphi$ can be turned into a standard $\AC$ formula by translating word equations and length constraints to rational constraints. 
Notice that, although quite powerful, 
extended $\AC$ still cannot express $\SL$ formulae such as $x = yy$, and does not cover practical properties such as, e.g., those in Example~\ref{ex:cacmSL} (where two conjuncts contain both $x$ and $y$).

\paragraph{Word equations to rational constraints}
For simplicity, assume that equations do not contain letters~$a \in \Sigma$.
This can be achieved by replacing every occurrence of a constraint $\symb$ by a fresh variable constrained by the regular language $\{\symb\}$. 
An equation~$x = x_{1} \concat \cdots \concat x_n$ without multiple occurrences of any variables is translated to a rational constraint $\aftrel(x,x_1,\ldots,x_n)$ with $\aftrel = (\bvarex {n+1},\states = \{q_0,\ldots,q_n\},\Delta,\init = q_0,\final = q_n)$. The transitions for $i\in [n]$ are   
$$\Delta(q_{i-1}) = (q_{i-1} \lor {q_i}) 
\land \bigwedge_{j\in[n]\setminus\{i\}} e^j \land 
\bigwedge_{v\in \bvarex{n+1}} (v^i \leftrightarrow v^0).
$$
and $\Delta(q_n) = \false$. That is, the symbol on the first track is copied to the $i$th track while all the other tracks read $\epsilon$. Negated word
equations can be translated to AFTs in a similar way. 

\paragraph{Length constraints to rational constraints}

The translation of length constraints to rational constraints is
similarly straightforward. Suppose an extended AC formula contains a
length constraint~$\varphi_{\text{Pres}}(|x_1|, \ldots, |x_k|)$, where
$\varphi_{\text{Pres}}$ is a Presburger formula over $k$
variables~$y_1, \ldots, y_k$ ranging over natural numbers.
It is a classical result that the
solution space of $\varphi_{\text{Pres}}$ forms a semi-linear
set~\cite{ginsburg1966}, i.e., can be represented as a finite union of
linear sets~$L_j = \{ \bar y_0 + \sum_{i=1}^m \lambda_i \bar y_i \mid
\lambda_1, \ldots, \lambda_m \in \nat \} \subseteq \nat^k$ with $\bar
y_0, \ldots \bar y_m \in \nat^k$. Every linear set~$L_j$ can directly
be translated to a succinct $k$-track AFT recognising the relation~$\{
(x_1, \ldots, x_k) \in (\Sigma^*)^k \mid (|x_1|, \ldots, |x_k|) \in
L_j \}$, and the union of AFTs be constructed as shown in
Section~\ref{sec:AFA-ops}, resulting in an
AFT~$\aftrel_{\varphi_{\text{Pres}}}(x_1, \ldots, x_k)$ that is
equivalent to $\varphi_{\text{Pres}}(|x_1|, \ldots, |x_k|)$.

\section{Rational Constraints with Synchronisation Parameters}
\label{sec:afa_parameters}

In order to simplify the decision procedure for $\SL$, which we will present in
Section~\ref{sec:sl}, we introduce an enriched syntax of rational constraints.
We will then extend the $\AC$ decision procedure from Section~\ref{sec:ac} to
the new type of constraints such that it can later be used as a subroutine in
our decision procedure of $\SL$. 
Before giving details, we will outline the main idea behind the
extension.

The $\AC$ decision procedure expects acyclicity, which prohibits formulae that
are, e.g., of the form $(\varphi(x)\land\varphi'(y))\land\psi(x,y)$.
Indeed, after replacing  the inner-most conjunction by an equivalent rational
constraint, the formula turns into the conjunction
$\aftrel_{\varphi\land\varphi'}(x,y)\land\aftrel_{\psi}(x,y)$, which is a
conjunction of the form $\aftrel(x,y) \land \aftrels(x,y)$.
In general, satisfiability of such conjunctions is not decidable, and they
cannot be expressed as a single AFT since synchronisation of $\epsilon$-moves on
multiple tracks is not always possible.
However, our example conjunction does not compose two arbitrary AFTs.
By its construction,  $\aftrel_{\varphi\land\varphi'}(x,y)$ actually consists of
two disjoint AFT parts.
Each of the parts constrains symbols read on one of the two tracks only and is
completely oblivious of the other part. 
Due to this, an AFT equivalent to
$\aftrel_{\varphi\land\varphi'}(x,y)\land\aftrel_{\psi}(x,y)$ can be constructed
(let us outline, without so far going into details, that the construction would
saturate $\epsilon$-moves for each track of $\aftrel_{\varphi\land\varphi'}$
separately).
Indeed, the original formula can also be rewritten as $\varphi(x)\land
(\varphi(y)\land\psi(x,y))$, which is $\AC$ and can be solved by the algorithm
of Section~\ref{sec:ac}.

The idea of exploiting the independence of tracks within a transducer can be
taken a step further.  The two independent parts do not have to be totally
oblivious of each other, as in the case of $\aftrel_{\varphi\land\varphi'}$
above, but can communicate in a certain limited way. 
To define the allowed form of communication and to make the independent
communicating parts syntactically explicit within string formulae, we will
introduce the notion of synchronisation parameters of AFTs.
We will then explain how formulae built from constraints with synchronisation
parameters can be transformed into a single rational constraint with parameters
by a simple adaptation of the $\AC$ algorithm,  and how the parameters can be
subsequently eliminated, leading to a single standard rational constraint. 


\begin{definition}[AFT with synchronisation parameters] An \emph{AFT with
parameters} $\vect s = s_1,\ldots,s_n$ is defined as a standard AFT $\aft =
(\bvar,\states,\transitions,\init,\final)$ with the difference that the initial
and the final formula can talk apart from states about so-called
\emph{synchronisation parameters} too.
That is, $\init,\final\subseteq\bofo{\states\cup\{\vect s\}}$ where $\init$ is
still positive on states and $\final$ is still negative on states, but the
synchronisation parameters can appear in $\init$ and $\final$ both positively as
well as negatively. 
The synchronisation parameters put an additional constraint on accepting runs.
A run $\run = \run_0 \ldots \run_m$ over a $k$-tuple of words $\vect w$ is
accepting only if there is a truth assignment $\pass:\{\vect s\}\rightarrow
\bool$ of parameters such that
$\pass\models\init$ and
$\pass\models\final$. 
We then say that $\vect w$ is \emph{accepted with the parameter assignment
$\pass$}.\end{definition}


String formulae can be built on top of AFTs with parameters in the same way as
before.
We write $\varphi[\vect s](\vect x)$ to denote a string formula that uses AFTs
with synchronisation parameters from $\vect s$ in its rational constraints.
Such a formula is interpreted over a union $\sass\cup\pass$ of an assignment
$\sass:\Var\varphi\rightarrow \power\bvar^*$ from string variables to strings, as
usual, and a parameter assignment $\pass:\{\vect s\}\rightarrow \bool$. 
An atomic constraint $\aftrel[\vect s](\vect x)$ is satisfied by
$\sass\cup\pass$, written $\sass\cup\pass\models\aftrel[\vect s](\vect x)$, if
$\aftrel$ accepts $(\sass(x_1),\ldots,\sass(x_{|\vect x|}))$ with the parameter
assignment $\pass$.
Atomic string constraints without parameters are satisfied by $\sass\cup\pass$
iff they are satisfied by $\sass$.
The satisfaction $\sass\cup\pass\models\varphi$ of a Boolean combination
$\varphi$ of atomic constraints  is then defined as usual.

Notice that within a non-trivial string formula, parameters may be shared among
AFTs of several rational constraints.
They then not only synchronise initial and final configuration of a single
transducer run, but provide the aforementioned limited way of communication
among AFTs of the rational constraints within the formula.


\begin{definition}[\AC with synchronisation parameters---\ACsp] The definition
of $\AC$ extends quite straightforwardly to rational constraints with parameters.
There is no other change in the definition except for allowing rational
constraints to use synchronisation parameters as defined above. \end{definition}


Notice that since we do not consider regular constraints with parameters,
constraints with parameters in $\ACsp$ formulae are never negated.

The synchronisation parameters allow for an easier transformation of string
formulae into $\AC$.
For instance, consider a formula of the form $\varphi(x,y) \land \psi(x,y)$
where one of the conjuncts, say $\varphi$, can be rewritten as $\varphi_1[\vect
s_1](x) \land \varphi_2[\vect s_2](y)$.
The whole formula can be written as $\varphi_1[\vect s_1](x) \land
(\varphi_2[\vect s_2](y) \land \psi(x,y))$, which falls into $\ACsp$. 
An example of such a formula $\varphi(x,y)$, commonly found in the benchmarks we
experimented with as presented later on, is a formula saying that $x\concat y$
belongs to a regular language, expressed by an AFA $\afa$.  
This can be easily expressed by a~conjunction $\aftrel_1[\vect s](x) \land
\aftrel_2[\vect s](y)$ of two unary rational constraints with parameters.
Intuitively, the AFTs $\aftrel_1$ and $\aftrel_2$ are two copies of $\afa$.
$\aftrel_1$ nondeterministicaly chooses a configuration where the prefix of
a~run of $\afa$ reading a word $x$ ends, accepts, and remembers the accepting
configuration in parameter values (it will have a parameter per state).
$\aftrel_2$ then reads the suffix of $x$, using the information contained in
parameter values to start from the configuration where $\aftrel_1$ ended.
We explain this construction in detail in Section~\ref{sec:sl}.

An $\ACsp$ formula $\varphi$ with parameters can be translated into a single,
parameter-free, rational constraint and then decided by an AFA language
emptiness check described in Section~\ref{sec:afaSolving}.
The translation is done in two steps: \begin{enumerate}

  \item A {\bf generalised $\AC$ algorithm} translates $\varphi(\vect x)$ to
  $\aftrel_\varphi[\bar s](\vect x)$.

  \item {\bf Parameter elimination} transforms $\aftrel_\varphi[\bar s](\vect
  x)$ to a normal rational constraint $\aftrel_\varphi'(\vect x)$.

\end{enumerate}

\paragraph{Generalised $\AC$ algorithm}

To enable eliminations of conjunctions and disjunctions from $\ACsp$ formulae,
just a small modification of the procedure from Section~\ref{sec:ac} is enough.  
The presence of parameters in the initial and final formulae does not require
any special treatment, except that, unlike for states (which are implicitly
renamed), it is important that sets of synchronisation parameters stay the same
even if they intersect, so that the synchronisation is preserved in the
resulting AFT.
That is, for $\square\in\{\land,\lor\}$,  $\aftrel_\varphi[\vect r](\vect x)$,
and $\aftrel_\psi[\vect s](\vect y)$, the constraint
$\aftrel_{\varphi\mathrel{\square}\psi}[\vect t](\vect z)$ is created in the
same way as described in Section~\ref{sec:ac}, the parameters within the initial
and the final formulae of the input AFTs are passed to the AFA construction
$\mathrel{\square}$ unchanged, and  $\{\vect t\} =
\{\vect r\} \cup \{\vect s\}$.

\begin{lemma} $\aftrel_\varphi[\vect r](\vect x) \mathrel{\square}
\aftrel_\psi[\vect s](\vect y)$ is equivalent to
$\aftrel_{\varphi\mathrel{\square}\psi}[\vect t](\vect z)$. \end{lemma}

\paragraph{Elimination of parameters}

The previous steps transform the formula into a single rational constraint with
synchronisation parameters. 
Within such a constraint, every parameter communicates one bit of information
between the initial and final configuration of a run.
The bit can be encoded by an additional automata state passed from a
configuration to a configuration via transitions through the entire run,
starting from an initial configuration where the parameter value is decided in
accordance with the initial formula, to the final configuration where it is
checked against the final formula. 
A technical complication, however, is that automata transitions are monotonic
(positive on states).
Hence, they cannot prevent arbitrary states from appearing in target
configurations even though their presence is not enforced by the source
configuration.
For instance, starting from a single state $q_1$ and executing a transition
$\Delta(q_1) = q_2$ can yield a configuration $q_2 \wedge q_3$.
The assignment of $0$ to a~parameter cannot therefore be passed through the run
in the form of absence of a single designated state as it can be overwritten
anywhere during the run. 

To circumvent the above, we use a so-called \emph{two rail encoding} of
parameter values: every parameter $s$ is encoded using a pair of value indicator
states, the positive value indicator $s^+$ and the negative value indicator
$s^-$. 
Addition of unnecessary states into target configurations during a~run then
cannot cause that a parameter silently changes its value.
One of the indicators can still get unnecessarily set, but the other indicator
will stay in the configuration too (states can be added into the configurations
reached, but cannot be removed).
The parameter value thus becomes ambiguous---both $s^-$ and $s^+$ are present.
The negative final formula can exclude all runs which arrive with
ambiguous parameters by enforcing that at least one of the indicators is
false. 

Formally, the parameter elimination replaces a constraint $\aftrel(\bar x)[\vect
s]$ with $\aftrel = (\bvarex{|\bar
x|},\controls,\transrel,\init,\final)$ and $|\bar s| = n$ by a parameter free constraint
$\aftrel'(\bar x)$ with $\aftrel' = (\bvarex{|\bar
x|},\controls',\transrel',\init',\final')$ where 
\begin{itemize} 

  \item  $\controls' = \controls\cup\{s_i^+,s_i^- \mid 1\leq i\leq n\}$ 
  (parameters are added to $Q$), and

  \item  $\transrel'=  \transrel\cup\{s_i^+\mapsto s_i^+,s_i^-\mapsto s_i^- \mid
  1\leq i\leq n\}$ (once active value indicators stay active).

  \item $\init' = \init^+ \land \mathit{Choose}$ where $\init^+$ is a positive
  formula that arises from $\init$ by replacing every negative occurrence of
  a~parameter $\neg s$ by a positive occurrence of its negative indicator $s^-$,
  and the positive formula $\mathit{Choose} = \bigwedge_{i=1}^n s_i^+\lor s_i^-$
  enforces that every parameter has a~value.

  \item $\final' = \final^- \land \mathit{Disambiguate}$ where $\final^-$ is a
  negative formula that arises from $\final$ by replacing every positive
  occurence of a~parameter $s$ by a negative occurrence of its negative indicator
  $\neg s^-$, and the negative formula $\mathit{Disambiguate} =
  \bigwedge_{i=1}^n \neg s_i^+\lor \neg s_i^-$ enforces that indicators
  determine parameter values unambiguously, i.e., at most one indicator per
  parameter~is~set.  

\end{itemize} 

\begin{lemma} $\exists \vect s:\aftrel(\bar x)[\vect s]$ is equivalent to
$\aftrel'(\bar x)$. \end{lemma}



\section{Deciding Straight-Line Formulae} \label{sec:sl}

Our algorithm solves string formulae using the DPLL(T) framework
\cite{tinelli_dpll_2004}\footnote{Also see \cite{KS08} for a gentle introduction
to DPLL(T).}, where $T$ is a sound and complete solver for $\AC$ and $\SL$. 
Loosely speaking, DPLL(T) can be construed as a collaboration between a
DPLL-based SAT-solver and theory solvers, wherein the input formula is viewed as
a Boolean formula by the SAT solver, checked for satisfiability by the
SAT-solver, and if satisfiable, theory solvers are invoked to check if the
Boolean assignment found by the SAT solver can in fact be realised in the
involved theories.  The details of the DPLL(T) framework are not so important
for our purpose.  However, the crucial point is that 
%
%
all queries that a DPLL(T) solver asks a T-theory solver are conjunctions from
the CNF of the input formula (or their parts), 
enabling us to concentrate on solving $\SL$ conjunctions only.

Our decision procedure for $\SL$ conjunctions transforms the input $\SL$
conjunction into an equisatisfiable $\ACsp$ formula, which is then decided as
discussed in Section~\ref{sec:afa_parameters}.
The rest of the section is thus devoted to a translation of a positive $\SL$
conjunction $\varphi$ to an $\ACsp$ formula.
The translation internally combines rational constraints and equations into a
more general kind of constraints in which rational relations are mixed with
concatenations and synchronisation parameters. 


\newcommand{\lb}{\mathtt{c}}
\newcommand{\rb}{\mathtt{d}}

\begin{example} \label{ex:brackets} As a running example for the section, we use
an $\SL$ conjunction that captures the essence of the vulnerability pattern from
Example~\ref{ex:cacm}: 
A sanitizer is applied on an input string to get rid of symbols $\lb$, replacing
them by $\rb$, hoping that this will prevent a dangerous situation which arises
when a symbol $\rb$ apears in a string somewhere behind $\lb$.
However, the dangerous situation will not be completely avoided since it is
forgotten that the sanitized string will be concatenated with another string
that can still contain $\lb$.\footnote{In reality, where one undesirably
concatenates a string $\mathtt{command('...}$ with some string $\mathtt{...');
attack();}$
the situation is, of course, more complex and
sanitization is more sophisticated. However, having a real-life example, such as those used in our
experiments, as a running example would be too complex to understand.}

To formalize the example, assume a bit-vector encoding of an alphabet $\Sigma$
which contains the symbols $\lb$ and $\rb$.
Assume that each $a \in\Sigma$ denotes the conjunction of (negated) bit
variables encoding it.
As our running example, we will then consider the formula $ \varphi : y =
\aftrel(x) \land z = x \concat y \land \afa(z)$.
The AFT $\aftrel = (\bvarex 2, Q = \{q\}, \Delta = \{q \mapsto q \land \neg
\rb^1 \land (\lb^1 \rightarrow \rb^2) \land
\bigwedge_{a\in\Sigma\setminus\{\lb\}} (a^1\leftrightarrow a^2)) \}, I = q, F =
\true)$ is a~sanitizer that produces $y$ by replacing all occurrences of $\lb$
in its input string $x$ by $\rb$, and it also makes sure that $x$ does not
include $\rb$.
The AFA $\afa = (\bvar, Q' = \{r_0,r_1,r_2\}, \Delta', I' = r_0, F' = \neg r_0
\land \neg r_1)$ where $\Delta'(r_0) = (r_0\land \neg \lb) \lor (r_1 \land
\lb)$, $\Delta'(r_1) = (r_1\land \neg \rb) \lor (r_2 \land \rb)$, and
$\Delta'(r_2) = \true $ is the specification.
It checks whether the opening symbol $\lb$ can be later followed by the closing
symbol $\rb$ in the string $z$.
The formula is satisfiable. \qed \end{example}


\begin{definition}[Mixed constraints] A \emph{mixed constraint} is of the form
$x= \aftrel[\bar s](y_1 \concat \cdots \concat y_n)$ where $\aftrel$ is a binary
AFT, with a concatenation of variables as the right-hand side argument, and
$\vect s$ is a~vector of synchronisation parameters.
Such constraint has the expected meaning: it is satisfied by the union
$\pass\cup\sass$ of an assignment $\sass$ to string variables and an assignment
$\pass$ to parameters iff
$(\sass(x),\sass(y_1)\concat\cdots\concat\sass(y_{n}))$ is accepted by
$\aftrel[\vect s]$ with the parameter assignment $\pass$. \end{definition}


All steps of our translation of the input SL formula $\varphi$ to an $\ACsp$
formula preserve the $\SL$ fragment, naturally generalised to mixed constraints
as follows.  


\begin{definition}[Generalised straight-line conjunction] A conjunction of
string constraints is defined to be generalised \emph{straight-line} if it can
be written as $\psi\land\bigwedge_{i=1}^m x_i = F_i$ where $\psi$ is a
conjunction over regular and negated regular constraints and each $F_i$ is
either of the form $y_1\concat\cdots\concat y_n$ or $\aft[\vect
s](y_1\concat\cdots \concat y_n)$ such that it does not contain variables
$x_{i},\ldots, x_m$. \end{definition}


For simplicity, we assume that $\varphi$ has gone through two preprocessing
steps.
First, all negations were eliminated by complementing regular constraints,
resulting in a purely positive conjunction. 
Second, all the---now only positive---regular constraints were replaced by
equivalent rational constraints. Particularly, a regular constraint $\afa(x)$ is
replaced by a rational constraint $x' = \aftrel'(x)$ where $x'$ is a fresh
variable and $\aftrel'$ is an AFT with $\rel({\aftrel'}) = \power\bvar^*\times
L(\afa)$.
The AFT $\aftrel'$ is created from $\afa$ by indexing all propositions in the
transition relation by the index $2$ of the second track.
It is not difficult to see that since $x'$ is fresh, the replacement preserves
$\SL$, and also satisfiability, since $P(x) \land \psi$ is equivalent to
$\exists x': x' = \aftrel(x) \land \psi$ for every $\psi$.


\begin{example} \label{ex:brackets0} In Example~\ref{ex:brackets}, the
preprocessing replaces the conjunct $\afa(z)$ by $z' = \aftrels(z)$ where
$\aftrels$ is the same as $\afa$, except occurrences of bit-vector
variables in $\Delta'$ are indexed by $2$ since $z$ will be read on its second
track.
We obtain $ \varphi'_0 : y = \aftrel(x) \land z = x \concat y \land z' =
\aftrels(z)$ where $z'$ is free. \qed \end{example}


Due to the preprocessing, we are starting with a formula $\varphi'_0$ in the
form of an $\SL$ conjunction of rational constraints and equations. The
translation to $\ACsp$ will be carried out in the following three steps, which
will be detailed in the rest of the section:\begin{enumerate}

  \item {\bf Substitution} transforms $\varphi'_0$ to a conjunction $\varphi_1$
  of mixed constraints.

  \item {\bf Splitting} transforms $\varphi_1$ to a conjunction $\varphi_2$
  of rational constraints with parameters. 

  \item {\bf Ordering} transforms $\varphi_2$ to an $\AC$ conjunction
  $\varphi_3$ with parameters.

\end{enumerate}

\paragraph{Substitution}

Equations in $\varphi'_0$ are combined with rational constraints into mixed
constraints by a~straightforward substitution. 
In one substitution step, a conjunction $x = y_1\concat\cdots\concat  y_n \land
\psi$ is replaced by $\psi[y_1\concat \cdots \concat y_n/x]$ where all
occurrences of $x$ are replaced by $y_1\concat \cdots \concat y_n$.
The substitution preserves the generalised straight-line fragment.

\begin{lemma} If $x = y_1\concat\cdots\concat  y_n \land \psi$ is $\SL$, then
$\psi[y_1\concat \cdots \concat y_n/x]$ is equisatisfiable and $\SL$.
\end{lemma}

The substitution steps are iterated eagerly in an arbitrary order until there
are no equations.
Every substitution step obviously decreases the number of equations, so the
iterative process terminates after a finitely many steps with an equation-free
$\SL$ conjunction of mixed constraints~$\varphi_1$. 


\begin{example} \label{ex:brackets1} The substitution eliminates the
equation $z = x\concat y$ in $\varphi'_0$ from Example~\ref{ex:brackets0},
transforming it to $\varphi_1: y = \aftrel(x)\land u = \aftrels(x\concat y)$.
\qed \end{example}

\paragraph{Splitting}

We will now explain how synchronisation parameters are used to eliminate
concatenation within mixed constraints. 
The operation of \emph{binary splitting} applied to an $\SL$ conjunction of
mixed constraints,
$\varphi: x = \aftrel(y_1\concat \cdots \concat y_m \
\concat \ z_1 \concat \cdots \concat z_n)[\vect s] \land \psi$,
where $\aftrel
= (\bvarex{2},\states,\transrel,\init,\final)$ and $\states =
\{q_1,\ldots,q_l\}$ splits the mixed constraint and substitutes $x$ by
a~concatenation of fresh variables $x_1\concat x_2$ in $\psi$.
That is, it outputs the conjunction
$\varphi':\zeta \land \psi[x_1\concat x_2/x]$
of mixed constraints, where the rational constraint was split into the
following conjunction $\zeta$ of two constraints:
$$\zeta: x_1 = \aftrel_1(y_1\concat\cdots\concat y_m)[\vect s , \vect t] \land
x_2 = \aftrel_{2}(z_1\concat\cdots\concat z_n)[\vect s , \vect t]$$
The vector $\vect t$ consists of $l$ fresh parameters, $x_1$ and $x_2$ are
fresh string variables, and each AFT with parameters $\aftrel_i =
(\bvarex{2},\states,\transrel,\init_i,\final_i),i\in\{1,2\}$, is derived
from $\aftrel$ by choosing initial/final formulae:
$$
I_1 = I, \quad
F_1 = \bigwedge_{i=1}^{l} q_i \rightarrow t_i,
\qquad
I_2 = \bigwedge_{i=1}^{l} t_i \rightarrow q_i,\quad
F_2 = F~.
$$
Intuitively, each run $\run$ of $\aftrel$ is split into a run $\run_1$ of
$\aftrel_1$, which corresponds to the first part of $\run$ in which $y_1\concat
\cdots \concat y_m$ is read along with a prefix $x_1$ of $x$, and a run $\run_2$
of $\aftrel_2$, which corresponds to the part of $\run$ in which $z_1\concat
\cdots \concat z_n$ is read along with the suffix $x_2$ of $x$.
Using the new synchronisation parameters $\vect t$, the formulae $F_1$ and $I_2$
ensure that the run $\run_1$ of $\aftrel_1$ must indeed start in the states in
which the run $\run_2$ of $\aftrel_2$ ended, that is, the original run $\run$ of
$\aftrel$ can be reconstructed by connecting $\run_1$ and $\run_2$.  
Every occurrence of $x$ in $\psi$ is replaced by the concatenation $x_1 \concat
x_2$.

\begin{lemma} In the above, $\varphi$ is equivalent to $\exists x_1 x_2 \vect t:
\varphi'$. \end{lemma}

The resulting formula $\varphi'$ is hence  equisatisfiable to the original
$\varphi$.
Moreover, $\varphi'$ is still generalised $\SL$---the two new constraints
defining $x_1$ and $x_2$ can be placed at the position of the original
constraint defining $x$ that was split, and the substitution $[x_1\concat
x_2/x]$ in the rest of the formula only applies to the right-hand sides of
constraints (since $x$ can be defined only once). 

\begin{lemma} If $\varphi$ is an $\SL$ conjunction of mixed constraints, then so
is $\varphi'$. \end{lemma}

Moreover, by applying binary splitting steps eagerly in an arbitrary order on
$\varphi_1$, we are guaranteed that all concatenations will be eliminated after
a finite number of steps, thus arriving at the $\SL$ conjunction of rational
constraints with parameters $\varphi_2$.
The termination argument relies on the straight-line restriction.
Although it cannot be simply said that every step reduces the number of
concatenations because the substitution $x_1\concat x_2$ introduces new ones, 
the new concatenations $x_1\concat x_2$ are introduced only into constraints
defining variables that are higher in the straight-line ordering than $x$.
It is therefore possible to define a well-founded (integer) measure on the
formulae that decreases with every application of the binary splitting steps. 

\begin{lemma}
  \label{lem:splittingTerm}
  All concatenations in the $\SL$ conjunction of mixed constraints
$\varphi_1$ will be eliminated after a finite number of binary splitting steps.
\end{lemma}
%
%
We note that our implementation actually uses a slightly more efficient $n$-ary
splitting instead of the described binary. 
It splits a mixed constraint in one step into the number of conjuncts equal to
the length of the concatenation in its right-hand side.
We present the simpler binary variant, which eventually
achieves the same effect.


\begin{example}\label{ex:brackets2}
The formula from Example~\ref{ex:brackets1} would be transformed into $\varphi_2: y = \aftrel(x) \land u_1 = \aftrels_1[\vect s](x) \land \aftrels_2[\vect s](y)$ where $\aftrels_1,\aftrels_2$ are as $\aftrels$ up to that $\aftrels_1$ has the final formula $I' \land \bigwedge_{i=0}^2(r_i \rightarrow s_0)$ and $\aftrels_2$ has the final formula $F' \land \bigwedge_{i=0}^2 (s_i \rightarrow r_i)$.  
Notice that $u_1 = \aftrels_1[\vect s](x) \land u_2 = \aftrels_2[\vect s](y)$ still enforce that $x\circ y$ has $\lb$ eventually followed by $\rb$. The parameters remember where $\aftrels_1$ ended its run and force $\aftrel_2$ to continue from the same state. \qed
\end{example} 

\paragraph{Reordering modulo associativity}

Substitution and splitting transform $\varphi_0$ to a straight-line conjunction $\varphi_2$ of rational constraints with parameters. 
Before delegating it to the $\ACsp$ formulae solver, 
it must be reorganized modulo associativity to achieve a structure satisfying the definition of $\AC$.
%
%
One way of achieving this is to order the formula into a 
conjunction $\bigwedge_{i=1}^m x_i = \aftrel[\vect s^i](y_i)$ satisfying the condition in the definition of $\SL$ (the definition of $\SL$ only requires that the formula \emph{can} be assumed). 
An simple way is discussed in \cite{LB16}. 
It consists of drawing the dependency graph of $\varphi$,
a directed graph with the variables $\Var\varphi$ as vertices which has an edge $x\rightarrow y$ if and only if $\varphi$ contains a conjunct $x = \aftrel(y)$. 
Due to the straight-line restriction, the graph must be acyclic. 
The ordering of variables can be then obtained as a topological sort of the graphs vertices, 
which is computable in linear time  (e.g. \cite{Cormen}, for instance by a depth-first traversal).
The final acyclic formula $\varphi_3$ then arises when letting $\bigwedge_{i=1}^m$ associate from the 
right:
$$\varphi_3: (x_{1} = \aftrel_{1}(y_1) \land ( x_{2} = \aftrel_{2}(y_2) \land (\ldots \land (x_{m-1} = \aftrel_{m-1}(y_{m-1}) \land x_m = \aftrel_m(y_m))\ldots))).$$
To see that $\varphi_3$ is indeed $\ACsp$, observe that every conjunctive sub-formula is of the form $(\bigwedge_{i<k} x_i = \aftrel_i(y_i)) \land x_k = \aftrel_k(y_k)$ where $x_k$ is by the definition of $\SL$ not present in the left conjunct. The left and right conjuncts can therefore share at most one variable, $y_k$. 

\begin{theorem}
The formula $\varphi_3$ obtained by substitution, splitting, and reordering from $\varphi_0$ is equisatisfiable and acyclic. 
\end{theorem}


\begin{example}
\label{ex:brackets3}
The $\ACsp$ formula
$\varphi_3:y = \aftrel(x) \land u_1 = \aftrels_1[\vect s](x)) \land \aftrels_2[\vect s](y)$
would be the final result of the $\SL$ to $\ACsp$ translation.
Let us use $\varphi_3$ to also briefly illustrate the decision procedure for $\ACsp$ of Section~\ref{sec:afa_parameters}.
The first step is the transformation to a single rational constraint with parameters by induction over  formula structure.
This will produce $\aftrel'[\vect s](x,y,z)$ with states and transitions consisting of those in $\aftrel$, $\aftrels_1$ with indexes of alphabet bits incremented by one ($y$, and $z$ are now not the first and the second, but the second and the third track), and a copy $\aftrels'_2$ of $\aftrels_2$ with states replaced by their primed variant (so that they are disjoint from that of $\aftrels_1$) and also incremented indexes of alphabet bits. The initial and final configuration will be the conjunctions of those of $\aftrel, \aftrels_1$ and $\aftrels_2'$.
The last step, eliminating of parameters, will lead to the addition of positive and negative indicator states for parameters $\vect s = s_1,s_2,s_3$ with the universal self-loops and the update of the initial and final formula as in Section~\ref{sec:afa_parameters}. The rest is solved by the emptiness check discussed in Section~\ref{sec:afaSolving}.
Notice the small size of the resulting AFT. Compared to the original formula from Example~\ref{ex:brackets}, it contains only one additional copy of $\afa$ (the $\aftrels_2')$, the six additional parameter indicator states with self-loops and the initial and final condition on the parameter indicators.
\qed
\end{example}

\paragraph{A note on the algorithm of \cite{LB16}}

\enlargethispage{1ex}
We will now comment on the differences of our algorithm for deciding $\SL$ from the earlier algorithm of \cite{LB16}. 
%
It combines reasoning on the level NFAs and nondeterministic transducers, utilising classical automata theoretic techniques, with a technique for eliminating concatenation by enumerative automata splitting.
It first turns and $\SL$ formula into a pure $\AC$ formula and then uses the $\AC$ decision procedure. 
An obvious advantage of our decision procedure described in Section~\ref{sec:ac} 
is the use of succinct AFA. 
As opposed to the worst case exponentially larger NFA, it produces an AFA of a linear size (unless the original formula contains negated regular constraints represented as general AFA. See Section~\ref{sec:ac} for a detailed discussion).
%
%
Let us also emphasize the advantages of our algorithm in the first phase, translation of $\SL$ to $\ACsp$.
%
%
Similarly as in the case of deciding $\AC$, 
the main advantage of our algorithm is that, while \cite{LB16} only works with NFTs, we propose ways of utilising the power of alternation and succinct transition encoding. 
%
%

We will illustrate the difference on an example.
%
%
The concatenation in the conjunction 
$x = y\concat z \land w = \aftrel(x)$ would in \cite{LB16} be done by enumerative splitting. 
It replaces the conjunction by the disjunction  
$\bigvee_{q\in Q}w_1 = \aftrel_q(y) \land w_2 = {_q\aftrel}(z)$. 
The $Q$ in the disjunction is the set of states of the (nondeterministic) transducer $\aftrel$, 
$\aftrel_q$ is the same as the NFT $\aftrel$ up to that the final state is $q$, 
and ${_q\aftrel}$ the same as $\aftrel$ up to that the initial state is $q$. 
Intuitively, the run of $\aftrel$ is explicitly separated into the part in which $y$ is read along the prefix $w_1$ of $w$, and the suffix in which $z$ is red along the suffix $w_2$ of $w$. 
The variable $w$ would be replaced by $w_1\concat w_2$ in the rest of the formula. 
The disjunction enumerates all admissible intermediate states $q
\in Q$ a run of $\aftrel$ can cross, and for each of them, it constructs two copies of $\aftrel$. This makes the cost of the transformation quadratic in the number of states of the NFT $\aftrel$.
A straightforward generalisation to our setting in which $\aftrel$ is an AFT is possible:
The disjunction would have to list, instead of possible intermediate states $q\in Q$, 
all possible intermediate configurations $C\subseteq Q$ a run of the AFA $\aftrel$ can cross, thus increasing the quadratic blow-up of the nondeterministic case to an exponential (due to the enumerative nature of splitting, the size is without any optimisation bounded by an exponential even from below). 
%



Our splitting algorithm utilises succinctness of alternation to reduce the cost of enumerative AFA splitting from exponential space (or quadratic in the case of NFAs) to linear. The smaller size of the resulting representation is payed for by a more complex alternating structure of the resulting rational constraints. 
The worst case complexity of the satisfiability procedure thus remains essentially the same. 
However, deferring most of the complexity to the last phase of the decision procedure, 
AFA emptiness checking, allows to circumvent the potential blow-up by means of modern model checking algorithms and heuristics and achieve much better scalability in practice.



\section{Model Checking for AFA Language Emptiness}
\label{sec:afaSolving}

In order to check unsatisfiability of a string formula using our
translation to AFTs, it is necessary to show that the resulting AFT
does not accept any word, i.e., that the recognised language is
empty. The constructed AFTs are succinct, but tend to be quite
complex: a na\"ive algorithm that would translate AFTs to NFAs using
an explicit subset construction, followed by systematic state-space
exploration, is therefore unlikely to scale to realistic string
problems.
We discuss how the problem of AFT emptiness can instead be reduced (in
linear time and space) to reachability in a~Boolean transition system,
in a way similar to \cite{fang-yu-circuits,arlen,Gange-mcstrings}.
Our translation is also inspired by the use of model checkers to
determinise NFAs in \cite{DBLP:conf/lpar/TabakovV05}, by a translation
to sequential circuits that corresponds to symbolic subset
construction. We use a similar implicit construction to map AFAs and
AFTs to NFAs.

As an efficiency aspect of the construction for AFAs, we observe that
it is enough to work with \emph{minimal} sets of states, thanks to the
monotonicity properties of AFAs (the fact that initial formulae and
transition formulae are positive in the state variables, and final
formulae are negative). This gives rise to three different versions: a
direct translation that does not enforce minimality at all; an
\emph{intensionally-minimal} translation that only considers minimal
sets by virtue of additional Boolean constraints; and a
\emph{deterministic} translation that resolves nondeterminism through
additional system inputs, but does not ensure fully-minimal state
sets.


\subsection{Direct Translation to Transition Systems}
\label{sec:afaSolving-direct}

To simplify the presentation of our translation to a Boolean
transition system, we focus on the case of AFAs $\afa =
(\bvarn,\states,\Delta,\init,\final)$ over a single track of
bit-vectors of length~$n+1$. The translation directly generalises to
$k$-track AFTs, and to AFTs with epsilon characters, by simply
choosing $n$ sufficiently large to cover the bits of all tracks.

We adopt a standard Boolean transition system view on the execution of
the AFA~$\afa$ (e.g., \cite{CGP99}). If $\afa$ has $m = |\states|$
automaton states, then $\afa$ can be interpreted as a (symbolically
described) transition system~$T^{\mathit{di}}_\afa = (\bool^m,
\mathit{Init}^{\mathit{di}}, \trel^{\mathit{di}})$. The transition
system has state space~$\bool^m$, i.e., a system state is a bit-vector
$\bar q = \langle q_0,\ldots,q_{m-1} \rangle$ of length~$m$ identifying
the active states in $\states$. The initial states of the system are
defined by $\mathit{Init}^{\mathit{di}}[\bar q] = I$, the same
positive Boolean formula as in $\afa$. The transition
relation~$\trel^{\mathit{di}}$ of the system is a Boolean formula over
two copies $\bar q, \bar q'$ of the state variables, encoding that for
each active pre-state $q_i$ in $\bar q$ the formula~$\Delta(q_i)$ has
to be satisfied by the post-state~$\bar q'$. Input variables $\bvarn =
\{ x_0, ..., x_n \}$ are existentially quantified in the transition
formula, expressing that all AFA transitions have to agree on the
letter to be read:
\begin{equation}
  \label{eq:treld}
  \trel^{\mathit{di}}[\bar q,\bar q'] ~=~
  \exists v_0, \ldots, v_n: \bigwedge_{i=0}^{m-1} q_i \to \Delta(q_i)[\bar q/\bar q']
\end{equation}

To examine emptiness of $\afa$, it has to be checked whether
$T^{\mathit{di}}_\afa$ can reach any state in the target
set~$\smash{\mathit{Final}^{\mathit{di}}}[\bar q] = F$, i.e., in the set
described by the negative final formula~$F$ of $\afa$. Since is
well-known that reachability in transition systems is a
PSPACE-complete problem~\cite{CGP99}, this directly establishes that
fragment~$\AC$ is in PSPACE (Corollary~\ref{cor:pspaceAC}).
\begin{lemma}
  The language~$L(\afa)$ recognised by the AFA~$\afa$ is empty if and
  only if $T^{\mathit{di}}_\afa$ cannot reach a configuration in
  $\mathit{Final}^{\mathit{di}}[\bar q]$.
\end{lemma}

In practice, this means that emptiness of $L(\afa)$ can be decided
using a wide range of readily available, highly optimised model
checkers from the hardware verification field, utilising methods such
as $k$-induction~\cite{DBLP:conf/fmcad/SheeranSS00}, Craig
interpolation~\cite{McMillan03}, or IC3/PDR~\cite{Bra12}. In our
implementation, we represent $T^{\mathit{di}}_\afa$ 
 in the
AIGER format~\cite{aiger}, and then apply
nuXmv~\cite{nuXmv} and ABC~\cite{abc}.

The encoding~$T^{\mathit{di}}_\afa$ leaves room for optimisation,
however, as it does not fully exploit the structure of AFAs and
introduces more transitions than strictly necessary. In
\eqref{eq:treld}, we can observe that if $\trel^{\mathit{di}}[\bar
q,\bar q']$ is satisfied for some~$\bar q, \bar q'$, then it will also
be satisfied for every post-state~$\bar q'' \succeq \bar q'$, writing
$\bar p \preceq \bar q$ for the point-wise order on bit-vectors~$\bar
p, \bar q \in \bool^m$ (i.e., $\bar p \preceq \bar q$ if $p_i$ implies
$q_i$ for every $i \in \{0, \ldots, m-1\}$). This is due to the
positiveness (or \emph{monotonicity}) of the transition
formulae~$\Delta(q_i)$. Similarly, since the initial formula~$I$ of an
AFA is positive, initially more states than necessary might be
activated. Because the final formula~$F$ is negative, and since
redundant active states can only impose additional restrictions on the
possible runs of an AFA, such redundant states can never lead to more
words being accepted.

More formally, we can observe that the transition
system~$T^{\mathit{di}}_\afa$ is
\emph{well-structured}~\cite{DBLP:conf/icalp/Finkel87}, which means
that the state space~$\bool^m$ can be equipped with a
well-quasi-order~$\leq$ such that whenever $\trel^{\mathit{di}}[\bar
q,\bar q']$ and $\bar q \leq \bar p$, then there is some state~$\bar
p'$ with $\bar q' \leq \bar p'$ and $\trel^{\mathit{di}}[\bar p,\bar
p']$.  In our case, $\leq$ is the inverse point-wise order~$\succeq$
on bit-vectors;\footnote{Since the state space~$\bool^m$ of
  $T^{\mathit{di}}_\afa$ is finite, the ``well-'' part is trivial.}
intuitively, deactivating AFA states can only enable more
transitions. Since the set~$\mathit{Final}^{\mathit{di}}[\bar q]$ is
upward-closed with respect to $\leq$ (downward-closed with respect to
$\preceq$), the theory on well-structured transition systems tells us
that it is enough to consider transitions to $\leq$-maximal states (or
$\preceq$-minimal states) of the transition system when checking
reachability of $\mathit{Final}^{\mathit{di}}[\bar q]$. In
forward-exploration of the reachable states of $T^{\mathit{di}}_\afa$,
the non-redundant states to be considered form an anti-chain.
%
%
This can be exploited by defining tailor-made exploration
algorithms~\cite{antichainsRaskin2010,IIC}, or, as done in the next
sections, by modifying the transition system to only include
non-redundant transitions.
%


\subsection{Intensionally-Minimal Translation}
\label{sec:afaSolving-min}

We introduce several restricted versions of the transition
system~$T^{\mathit{di}}_\afa$, by removing transitions to non-minimal
states. The strongest transition system~$\smash{T^{\text{min}}_\afa =
  (\bool^m, \mathit{Init}^{\text{min}}, \trel^{\text{min}})}$ obtained
in this way can abstractly be defined as:
\begin{align}
  \label{eq:initMin}
  \mathit{Init}^{\text{min}}[\bar q] &~=~
  \mathit{Init}^{\mathit{di}}[\bar q] \wedge
  \forall \bar p \prec \bar q.\; \neg \mathit{Init}^{\mathit{di}}[\bar p]
  \\[-1.5mm]
  \label{eq:transMin}
  \trel^{\text{min}}[\bar q, \bar q'] &~=~
  \trel^{\mathit{di}}[\bar q, \bar q'] \wedge
  \forall \bar p \prec \bar q'.\; \neg \trel^{\mathit{di}}[\bar q, \bar
  p]
\end{align}
That means, $\mathit{Init}^{\text{min}}$ and $\trel^{\text{min}}$ are
defined to only retain the $\preceq$-minimal states.  Computing
$\mathit{Init}^{\text{min}}$ and $\trel^{\text{min}}$ corresponds to
the logical problem of
\emph{circumscription}~\cite{DBLP:journals/ai/McCarthy80}, i.e., the
computation of the set of minimal models of a formula.
Circumscription is in general computationally hard, and its precise
complexity still open in many cases; in \eqref{eq:initMin} and
\eqref{eq:transMin}, note that eliminating the universal quantifiers
(as well as the universal quantifiers introduced by negation of
$\trel^{\mathit{di}}$) might lead to an exponential increase in
formula size, so that $T^{\text{min}}_\afa$ does not directly appear
useful as input to a model checker.

We can derive a more practical, but weaker system $T^{\text{im}}_\afa
= (\bool^m, \mathit{Init}^{\text{im}}, \trel^{\text{im}})$ by only
minimising post-states in $\trel^{\text{im}}$ with respect to the same
input letter~$\bvarn$:
\begin{align*}
  \mathit{Init}^{\text{im}}[\bar q] &~=~
  \mathit{Init}^{\text{min}}[\bar q]
  \\
  \trel^{\text{im}}[\bar q, \bar q'] &~=~
  \exists\, \bvarn.\;
  \Big(
  \trel[\bar q, \bar q', \bvarn]
  \wedge
  \forall \bar p \prec \bar q'.\;
  \neg \trel[\bar q, \bar p, \bvarn]
  \Big)
  \\
  \text{with}\quad
  \trel[\bar q, \bar q', \bvarn] &~=~
  \bigwedge_{i=0}^{m-1} q_i \to \Delta(q_i)[\bar q/\bar q']
\end{align*}
The formulae still contain universal quantifiers~$\forall \bar p$, but
it turns out that the quantifiers can now be eliminated with only
polynomial effort, due to the fact that $\bar p$ only occurs
negatively in the scope of the quantifier. Indeed, whenever
$\varphi[\bar q]$ is a formula that is positive in $\bar q$, and
$\varphi[\bar q]$ holds for assignments~$\bar q_1, \bar q_3 \in
\bool^m$ with $\bar q_1 \preceq \bar q_3$, then $\varphi[\bar q]$ will
also hold for any assignment~$\bar q_2 \in \bool^m$ with $\bar q_1
\preceq \bar q_2 \preceq \bar q_3$ due to monotonicity. This implies
that a satisfying assignment~$\bar q_1 \in \bool^m$ is
$\preceq$-minimal if no single bit in $\bar q_1$ can be
switched from $1$ to $0$ without violating $\varphi[\bar
q]$. More formally,
$
  \varphi[\bar q]
  \wedge \neg \exists \bar p \prec \bar q.~ \varphi[\bar p]$
  is equivalent to
  $\varphi[\bar q] \wedge
  \bigwedge_{i=0}^{m-1} q_i \to \neg\varphi[\bar q][q_i/\false]$,
where we write $\varphi[q_i/\false]$ for the result of substituting
$q_i$ with $\false$ in $\varphi$.

The corresponding, purely existential representation of
$\mathit{Init}^{\text{im}}$ and $\trel^{\text{im}}$ is:
\begin{align}
  \label{eq:initIm2}
  \mathit{Init}^{\text{im}}[\bar q] &~\equiv~
  \mathit{Init}^{\mathit{di}}[\bar q] \wedge
  \bigwedge_{i=0}^{m-1}
  q_i \to \neg \mathit{Init}^{\mathit{di}}[\bar q][q_i/\false]
  \\
  \label{eq:transIm2}
  \trel^{\text{im}}[\bar q, \bar q'] &~\equiv~
  \exists\, \bvarn.\;
  \Big(
  \trel[\bar q, \bar q', \bvarn]
  \wedge
  \bigwedge_{i=0}^{m-1}
  q'_i \to 
  \neg \trel[\bar q, \bar q', \bvarn][q'_i/\false]
  \Big)
\end{align}
The representation is quadratic in size of the original
formulae~$\mathit{Init}^{\mathit{di}}, \trel^{\mathit{di}}$, but the
formulae can in practice be reduced drastically by sharing of common
sub-formulae, since the $m$ copies of
$\mathit{Init}^{\mathit{di}}[\bar q][q_i/\false]$ and
$\trel[\bar q, \bar q', \bvarn][q'_i/\false]$ tend to be almost
identical.

\begin{lemma}
  The following statements are equivalent:
  \begin{enumerate}
  \item $T^{\mathit{di}}_\afa$ can reach a configuration in
    $\mathit{Final}^{\mathit{di}}[\bar q]$;
  \item $T^{\mathit{min}}_\afa$ can reach a configuration in
    $\mathit{Final}^{\mathit{di}}[\bar q]$;
  \item $T^{\mathit{im}}_\afa$ can reach a configuration in
    $\mathit{Final}^{\mathit{di}}[\bar q]$.
  \end{enumerate}
\end{lemma}

\begin{example}
  \label{ex:encodingIm}
  To illustrate the $T^{\mathit{im}}_\afa$ encoding, we consider an
  AFA $\afa$ that accepts the language $\{ xwy \mid |xwy| = 2k, k \geq
  1, x \in \{a,b\}, y \in \{c,d\} \}$ using the encoding of the
  alphabet $\Sigma = \{ a,b,c,d \}$ from Example~\ref{ex:alphabet}. We
  let $\afa = ( \{v_0, v_1\}, \{q_0,q_1,q_2,q_3,q_4 \}, \Delta, \init,
  \final)$ where $\init = q_0$, $\final = \neg q_0 \wedge \neg q_1
  \wedge \neg q_3 $ (i.e., the accepting states are $q_2$ and $q_4$),
  and $\Delta$ is defined as
  $\Delta(q_0) = \neg v_1 \wedge q_1 \wedge q_3$,
  $\Delta(q_1) = q_2$,
  $\Delta(q_2) = q_1$,
  $\Delta(q_3) = q_3 ~\vee~ (v_1 \wedge q_4)$, and
  $\Delta(q_4) = \false$.
  
  The direct transition system representation is $T^{\mathit{di}}_\afa
  = (\bool^5, \mathit{Init}^{\mathit{di}}, \trel^{\mathit{di}})$,
  defined by:
  \begin{equation*}
    \mathit{Init}^{\mathit{di}}[\bar q] ~=~ q_0,
    \qquad
    \trel^{\mathit{di}}[\bar q, \bar q'] ~=~
    \exists v_0, v_1.~
    \underbrace{
      \left(
    \begin{array}{@{}l@{}}
      (q_0 \to \neg v_1 \wedge q'_1 \wedge q'_3) \wedge\mbox{}
      \\
      (q_1 \to q'_2) \wedge\mbox{}
      \\
      (q_2 \to q'_1) \wedge\mbox{}
      \\
      (q_3 \to q'_3 ~\vee~ (v_1 \wedge q'_4)) \wedge\mbox{}
      \\
      (q_4 \to \false)
    \end{array}
  \right)}_{\trel[\bar q, \bar q', \bvarn]}
  \end{equation*}
  The intensionally-minimal translation $T^{\mathit{im}}_\afa$ can be
  derived from $T^{\mathit{di}}_\afa$ by conjoining the restrictions
  in \eqref{eq:initIm2} and \eqref{eq:transIm2}
  ($\trel^{\mathit{im}}[\bar q, \bar q']$ is shown in simplified form
  for sake of presentation):
  \begin{align*}
    \mathit{Init}^{\mathit{im}}[\bar q]
    &~=~
      q_0 \wedge
      (q_0 \to \neg\false) \wedge
      \bigwedge_{i=1}^{4} (q_i \to \neg q_0)
    ~\equiv~
      q_0 \wedge \neg q_1 \wedge \neg q_2 \wedge \neg q_3 \wedge \neg q_4
    \\
    \trel^{\mathit{im}}[\bar q, \bar q']
    &~\equiv~
    \exists v_0, v_1.~
      \left(
      \begin{array}{@{}l@{}}
        \trel[\bar q, \bar q', \bvarn] \wedge
        \neg q'_0 \wedge
        (q'_1 \to q_0 \vee q_2) \wedge
        (q'_2 \to q_1 ) \wedge\mbox{} \\
        (q'_3 \to q_0 \vee
        (q_3 \wedge \neg (v_1 \wedge q'_4))) \wedge
        (q'_4 \to q_3 \wedge \neg q'_3)
      \end{array}
      \right)
  \end{align*}

  \vspace*{-4ex}
  \qed

  %
\end{example}


\subsection{Deterministic Translation}
\label{sec:afaSolving-det}

We introduce a further encoding of $\afa$ as a transition system that
is more compact than \eqref{eq:initIm2}, \eqref{eq:transIm2}, but does
not always ensure fully-minimal state sets. The main idea of the
encoding is that a~conjunctive transition formula~$\Delta(q_1) = q_2
\wedge q_3$, assuming that $q_2, q_3$ do not occur in any other
transition formula~$\Delta(q_i)$, can be interpreted as a set of
deterministic updates~$q'_2 = q_1; q'_3 = q_1$. For state variables
that occur in multiple transition formulae, the right-hand side of the
update turns into a disjunction. Disjunctions in transition formulae
represent nondeterministic updates that can be resolved using
additional Boolean flags. The resulting transition system is
deterministic, as transitions are uniquely determined
by the pre-state and variables representing system inputs.

\begin{example}
  We illustrate the encoding $T^{\mathit{det}}_\afa = (\bool^m,
  \mathit{Init}^{\mathit{det}}, \trel^{\mathit{det}})$ using the AFA
  from Example~\ref{ex:encodingIm}. While the initial
  states~$\smash{\mathit{Init}^{\mathit{det}}}[\bar q]$ coincide with
  $\smash{\mathit{Init}^{\mathit{im}}}[\bar q]$ in
  Example~\ref{ex:encodingIm}, the transition
  relation~$\trel^{\mathit{det}}[\bar q, \bar q']$ now consists of two
  parts: a deterministic assignment of the post-state~$\bar q'$ in
  terms of the pre-state~$\bar q$, together with an auxiliary
  variable~$h_3$ that determines which branch of $\Delta(q_3)$ is
  taken; and a conjunct that ensures that value of $h_3$ is consistent
  with the inputs~$\bvarn$. The resulting $\trel^{\mathit{det}}[\bar
  q, \bar q']$ is (in this example) equivalent
  to $\trel^{\mathit{im}}[\bar q, \bar q']$:
  \begin{align*}
    \mathit{Init}^{\mathit{det}}[\bar q]
    &~=~
      q_0 \wedge \neg q_1 \wedge \neg q_2 \wedge \neg q_3 \wedge \neg q_4
    \\
    \trel^{\mathit{det}}[\bar q, \bar q']
    &~\equiv~
    \exists h_3.\;
    \left(
    \begin{array}{@{}l@{}}
      (q'_0 \leftrightarrow \false) \wedge\mbox{}\\
      (q'_1 \leftrightarrow q_0 \vee q_2) \wedge\mbox{}\\
      (q'_2 \leftrightarrow q_1) \wedge\mbox{}\\
      (q'_3 \leftrightarrow q_0 \vee q_3 \wedge h_{3}) \wedge\mbox{}\\
      (q'_4 \leftrightarrow q_3 \wedge \neg h_{3})
    \end{array}
    \right)
    \wedge
    \exists v_0, v_1.\;
    \left(
    \begin{array}{@{}l@{}}
      (q_0 \to \neg v_1) \wedge\mbox{}\\
      (q_3 \wedge \neg h_3 \to v_1) \wedge\mbox{}\\
      (q_4 \to \false)
    \end{array}
    \right)
  \end{align*}

  \vspace*{-5ex}
  \qed
\end{example}

To define the encoding formally, we make the simplifying assumption
that there is a unique initial state~$q_0$, i.e., $I = q_0$, and that
all transition formulae~$\Delta(q_i)$ are in negation normal form
(i.e., in particular state variables in $\Delta(q_i)$ do not occur
underneath negation). Both assumption can be established by simple
transformation of $\afa$. The transition system~$T^{\mathit{det}}_\afa
= (\bool^m, \mathit{Init}^{\mathit{det}}, \trel^{\mathit{det}})$
is:
\begin{align*}
  \mathit{Init}^{\text{det}}[\bar q] &~=~
  q_0 \wedge \bigwedge_{i =1}^{m-1} \neg q_i
  \\
  \trel^{\text{det}}[\bar q, \bar q'] &~=~
  \exists H.~ \left(
  \Big(\bigwedge_{i=0}^{m-1} q'_i \leftrightarrow \mathit{NewState}(q_i)\Big)
  \wedge
  \exists \bvarn.~
  \Big(\bigwedge_{i=0}^{m-1} q_i \to \mathit{InputInv}(\Delta(q_i), i)\Big)
  \right)
\end{align*}

The transition relation~$\trel^{\text{det}}$ consists of two main
parts: the state updates, which assert that every post-state
variable~$q'_i$ is set to an update formula~$\mathit{NewState}(q_i)$;
and an input invariant asserting that the letters that are read are consistent
with the transition taken. To determinise disjunctions in transition
formulae~$\Delta(q_i)$, a set~$H$ of additional Boolean
variables~$h_l$ (uniquely indexed by a position sequence~$l \in \Z^*$)
is introduced, and existentially quantified in $\trel^{\text{det}}$.

The update formulae~$\mathit{NewState}(q_i)$ are defined as a
disjunction of assignments extracted from the transition
formulae~$\Delta(q_j)$,
\begin{equation*}
  \mathit{NewState}(q_i) ~=~
  \bigvee \{ \varphi \mid
  \text{there is~} j \in \{0, \ldots, m-1\} \text{~such that~}
  \langle q_i, \varphi \rangle \in \mathit{StateAsgn}(\Delta(q_j), j, q_j) \}
\end{equation*}
where each $\mathit{StateAsgn}(\Delta(q_j), j, q_j)$ represents the
set of asserted state variables~$q_i$ in $\Delta(q_j)$, together with
guards~$\varphi$ for the case that $q_i$ occurs underneath
disjunctions. The set is recursively defined (on formulae in NNF) as
follows:
\begin{align*}
  \mathit{StateAsgn}(\varphi_1 \wedge \varphi_2, l, g) &~=~
  \mathit{StateAsgn}(\varphi_1, l, g) \cup
  \mathit{StateAsgn}(\varphi_2, l, g)
  \\
  \mathit{StateAsgn}(\varphi_1 \vee \varphi_2, l, g) &~=~
  \mathit{StateAsgn}(\varphi_1,\, l.1,\, g \wedge h_l) \cup
  \mathit{StateAsgn}(\varphi_2,\, l.2,\, g \wedge \neg h_l)
  \\
  \mathit{StateAsgn}(q_i, l, g) &~=~ \{ \langle q_i, g \rangle \}
  \\
  \mathit{StateAsgn}(\phi, l, g) &~=~ \emptyset
  \qquad\qquad (\text{for any other~} \phi)~.
\end{align*}
In particular, the case for disjunctions~$\varphi_1 \vee \varphi_2$
introduces a fresh variable~$h_l \in H$ (indexed by the position~$l$
of the disjunction) that controls which branch is taken. Input
variables~$v_i \in \bvarn$ are ignored in the updates.

The input invariants~$\mathit{InputInv}(\Delta(q_i), i)$ are similarly
defined recursively, and include the same auxiliary variables $h_l \in
H$, but ensure input consistency:
\begin{align*}
  \mathit{InputInv}(\varphi_1 \wedge \varphi_2, l) &~=~
  \mathit{InputInv}(\varphi_1, l) \wedge
  \mathit{InputInv}(\varphi_2, l)
  \\[-1mm]
  \mathit{InputInv}(\varphi_1 \vee \varphi_2, l) &~=~
  \big(h_l \to \mathit{InputInv}(\varphi_1, l.1)\big) \wedge
  \big(\neg h_l \to \mathit{InputInv}(\varphi_2, l.2)\big)
  \\[-1mm]
  &\hspace*{-28mm}
  \mathit{InputInv}(v_i, l) ~=~ v_i,
  \hspace*{3mm}
  \mathit{InputInv}(\neg v_i, l) ~=~ \neg v_i,
  \hspace*{3mm}
  \mathit{InputInv}(q_i, l) ~=~ \true,
  \hspace*{3mm}
  \mathit{InputInv}(\phi, l) ~=~ \phi~.
\end{align*}

\vspace*{-6mm}
\section{Implementation and Experiments} \label{sec:experiments}

We have implemented our method for deciding conjunctive AC and SL formulae as a
solver called \strsolver\ (\textbf{S}tring \textbf{LO}gic \textbf{TH}eory solver), extending the Princess
SMT solver~\cite{princess08}.  
The solver \strsolver\ can be obtained from
\url{https://github.com/uuverifiers/sloth/wiki}.
Hence, Princess provides us with infrastructure
such as an implementation of DPLL(T) or facilities for reading input formulae in
the \smtlib\ format \cite{smtlibv2:10}. Like Princess, \strsolver\ was implemented in Scala.
We present results from several settings of our tool featuring different optimizations.
\begin{description}
[
itemsep      =0pt,
leftmargin   =15pt,
]

\item{\strsolver-1}
The basic version of \strsolver, denoted as \strsolver-1 below, uses the direct
translation of the AFA emptiness problem to checking reachability in transition
systems described in Section~\ref{sec:afaSolving-direct}. Then, it employs the
nuXmv model checker \cite{nuXmv} to solve the reachability problem via the IC3
algorithm \cite{Bra12}, based on property-directed state space approximation.
Further, we have implemented five optimizations/variants of the basic solver:
four of them are described below, the last one at the end of the section.


\item{\strsolver-2}
Our first optimization, implemented in \strsolver-2, is rather simple: We assume
working with strings over an alphabet $\Sigma$ and look for equations of the
form $x = a_0 \concat y_1 \concat a_1 \ldots \concat y_n \concat a_n$ where $n
\geq 1$, $\forall 0 \leq i \leq n: a_i \in \Sigma^*$ (i.e., $a_i$ are constant
strings), and, for every $1 \leq j \leq n$, $y_j$ is a~free string variable not
used in any other constraint. The optimization replaces such constraints by
a~regular constraint $(a_0 \concat \Sigma^* \concat a_1 \ldots \concat \Sigma^*
\concat a_n)(x)$. This step allows us to avoid many split operations. The
optimization is motivated by a~frequent appearance of constraints of the given
kind in some of the considered benchmarks. As shown by our experimental
results below, the optimization yields very
significant savings in practice,  despite of its simplicity.

\item{\strsolver-3} Our second optimization, implemented in \strsolver-3,
replaces the use of nuXmv and IC3 in \strsolver-2\ by our own, rather simple
model checker working directly on the generated AFA. In particular, our model
checker is used whenever no split operation is needed after the preprocessing
proposed in our first optimization. It works explicitly with sets of conjunctive
state formulae representing the configurations reached. The initial formula and
transition formulae are first converted to DNF using the Tseytin procedure. Then
a SAT solver---in particular, sat4j \cite{DBLP:journals/jsat/BerreP10}---is used
to generate new reachable configurations and to check the final condition. Our
experimental results show that using this simple model checking approach can win
over the advanced IC3 algorithm on formulae without splitting.

\item{\strsolver-4}
Our further optimization, \strsolver-4, optimizes \strsolver-3 by employing the
intensionally minimal successor computation of Section~\ref{sec:afaSolving-min}
within the IC3-based model checking of nuXmv. 
\item{\strsolver-5}
Finally, \strsolver-5\ modifies
\strsolver-4\ by replacing the use of nuXmv with the property directed
reachability (i.e., IC3) implementation in the ABC tool \cite{abc}.
\end{description}

We present data on two benchmark groups (each consisting of two benchmark sets) that demonstrate two points.
First, the main strength of our tool is shown on solving complex combinations of
transducer and concatenation constraints (generated from program code similar to that of Example~\ref{ex:cacm}) 
that are beyond capabilities of any other solver.
Second, we show that our tool is competitive also on simpler examples that can
be handled by other tools (smaller constraints with less intertwined and general combinations of rational and concatenation constraints). 
%
%
All the benchmarks fall within the decidable straight-line fragment (possibly
extended with the restricted length constraints).
All experiments were 
executed on a computer with Intel Xeon E5-2630v2 CPU @ 2.60 GHz and 32 GiB RAM.


\paragraph{Complex combinations of concatenation and rational constraints.}
The first set of our benchmarks consisted of 10 formulae (5 sat and 5 unsat)
derived manually from the PHP programs available from the web page of the
\stranger\ tool \cite{Stranger}. The property checked was absence of the
vulnerability pattern \verb|.*<script.*| in the output of the programs. The
formulae contain 7--42 variables (average 21) and 7--38 atomic constraints
(average 18). Apart from the Boolean connectives $\wedge$ and $\vee$, they use
regular constraints, concatenation, the \texttt{str.replaceall} operation, and
several special-purpose transducers encoding various PHP functions used in the
programs (e.g., \texttt{addslashes}, \texttt{trim}, etc.).

\begin{wraptable}{r}{80mm}
\begin{center}
  \vspace*{-5mm}
  \caption {PHP benchmarks from the web of \stranger.\vspace*{-4mm}}
  \label{tab:stranger:php}
  \begin{tabular}{| l | r | r | r | r | r |} \hline
    Program & \#sat (sec) & \#unsat (sec) & \#mo & \#win +/- \\ \hline
    \strsolver-1 & 4 (178) & 5 (6,989) & 1 & 1/0 \\ \hline
    \strsolver-2 & 4 (83) & 5 (5,478) & 1 & 0/2 \\ \hline
    \strsolver-3 & 4 (72) & 5 (3,673) & 1 & 1/2 \\ \hline
    \strsolver-4 & 4 (93) & 4 (6,168) & 2 & 0/0 \\ \hline
    \strsolver-5 & 4 (324) & 4 (4,409) & 2 & 2/1 \\ \hline
  \end{tabular}
  \vspace*{-4mm}
\end{center}
\end{wraptable}
Results of running the different versions of \strsolver\ on the formulae are
shown in Table~\ref{tab:stranger:php}. Apart from the \strsolver\ version used,
the different columns show numbers of solved sat/unsat formulae (together with
the time used), numbers of out-of-memory runs (``mo''), as well as numbers of
sat/unsat instances for which the particular \strsolver\ version provided the
best result (``win +/-''). We can see that \strsolver\ was able to solve 9 out
of the 10 formulae, and that each of its versions---apart from
\strsolver-4---provided the best result in at least some case.
%


Our second benchmark consists of 8 challenging formulae taken from the paper
\cite{Kern14} providing an overview of XSS vulnerabilities in JavaScript
programs (including the motivating example from the introduction). \\
\begin{wraptable}{r}{71mm}
\begin{center}
\vspace*{-5mm}
  \caption {Benchmarks from \cite{Kern14}.\vspace*{-4mm}}
  \label{tab:popl}
  \begin{tabular}{| l | r | r | r | r |} \hline
    Solver & \#sat (sec) & \#unsat (sec) & \#win +/- \\ \hline
    \strsolver-1 & 4 (458) & 4 (583) & 0/2 \\ \hline
    \strsolver-2 & 4 (483) & 4 (585) & 0/1 \\ \hline
    \strsolver-3 & 4 (508) & 4 (907) & 2/1 \\ \hline
    \strsolver-4 & 4 (445) & 4 (1,024) & 1/0 \\ \hline
    \strsolver-5 & 4 (568) & 4 (824) & 1/0 \\ \hline
  \end{tabular}
  \vspace*{-5mm}
\end{center}
\end{wraptable}
The formulae
contain 9--12 variables (average 9.75) and 9--13 atomic constraints (average
10.5). Apart from conjunctions, they use regular constraints, concatenation,
\texttt{str.replaceall}, and again several special-purpose transducers encoding
various JavaScript functions (e.g., \texttt{htmlescape}, \texttt{escapeString},
etc.).
The results of our experiments are shown in Table~\ref{tab:popl}. The meaning of
the columns is the same as in Table~\ref{tab:stranger:php} except that we drop the
out-of-memory column since \strsolver\ could handle all the formulae---which we
consider to be an excellent result.

These results are the highlight of our experiments, taking into account that we are not
aware of any other tool capable of handling the logic fragment
used in the formulae.\footnote{We tried to replace the special-purpose
transducers by a sequence of \texttt{str.replaceall} operations in order to
match the syntactic fragment of the \sthreep\ solver \cite{joxan-cav16}. 
However,
neither \strsolver\ nor \sthreep\ could handle the modified formulae. We have
not experimented with other semi-decision procedures, such as those implemented
within \stranger\ or \slog\ \cite{fang-yu-circuits}, since they are indeed a different kind
of tool, 
and, moreover, often are not able to process input in the \smtlib\ format, which
would complicate the experiments.}
%



%
%


\paragraph{A Comparison with other tools on simpler benchmarks.}
Our next benchmark consisted of 3,392 formulae provided to us by the authors of
the \stranger\ tool. These formulae were derived by \stranger\ from real web
applications analyzed for security; to enable other tools to handle
the benchmarks, in the benchmarks the
\texttt{str.replaceall} operation was approximated by \texttt{str.replace}.
Apart from the $\wedge$ and
$\vee$ connectives, the formulae use regular constraints, concatenation, and the
\texttt{str.replace} operation. They contain 1--211 string variables (on average
6.5) and 1--182 atomic formulae (on average 5.8). 
Importantly, the use of concatenation is much less intertwined with  \texttt{str.replace} than it is with rational constraints in benchmarks from Tables \ref{tab:stranger:php} and \ref{tab:popl} (only about 120 from the 3,392 examples contain \texttt{str.replace}). 
Results of experiments on
this benchmark are shown in Table~\ref{tab:cvc4}.
In the table, we compare the different versions of our \strsolver, the \sthreep\
solver, and the \cvc\ string solver \cite{cvc4}.\footnote{The \sthreep\ solver
and \cvc\ solvers are taken as two representatives of semi-decision procedures
for the given fragment with input from \smtlib.} The meaning of the columns is
the same as in the previous tables, except that we now specify
both the number of
time-outs (for a~time-out of 5 minutes) and out-of-memory runs (``to/mo'').

\begin{wraptable}{r}{90mm}
\begin{center}
  \vspace*{-5mm}
  \caption {Benchmarks from \stranger\ with \texttt{str.replace}.\vspace*{-4mm}}
  \label{tab:cvc4}
  \begin{tabular}{| l | r | r | r | r | r |} \hline
    Solver & \#sat (sec) & \#unsat (sec) & \#to/mo & \#win +/- \\ \hline
    \strsolver-1 & 1,200 (19,133) & 2,079 (3,276) & 105/8 & 30/43 \\ \hline
    \strsolver-2 & 1,211 (13,120) & 2,079 (3,338) & 97/5 & 19/0 \\ \hline
    \strsolver-3 & 1,290 (6,619) & 2,082 (1,012) & 14/6 & 263/592 \\ \hline
    \strsolver-4 & 1,288 (6,240) & 2,082 (1,030) & 17/5 & 230/327 \\ \hline
    \strsolver-5 & 1,291 (6,460) & 2,082 (953) & 14/5 & 768/1,120 \\ \hline
    \cvc\ & 1,297 (857) & 2,082 (265) & 13/0 & -- \\ \hline
    S3P\ & 1,291 (171) & 2,078 (56) & 13/0 & -- \\ \hline
  \end{tabular}
  \vspace*{-4mm}
\end{center}
\end{wraptable}

From the results, we can see that \cvc\ is winning, 
but (1)~unlike \strsolver, it is a semi-decision procedure only, and (2)~the formulae of this benchmark are
much simpler than in the previous benchmarks (from the point of view of the
operations used), 
and hence the power of \strsolver\ cannot really manifest.

Despite that, our solver succeeds in almost the same number of examples as \cvc,
and it is reasonably efficient. 
Moreover, a closer analysis of the results reveals that our
solver won in 16 sat and 3 unsat instances. Compared with \sthreep, \strsolver\
won in 22 sat and 4 unsat instances (plus \sthreep\ provided 8 unknown and 1
wrong answer and also crashed once). This shows that \strsolver\ can compete
with semi-decision procedures at least in some cases even on a~still quite
simple fragment of the logic it supports. 


\begin{wraptable}{r}{85mm}
\begin{center}
  \vspace*{-5mm}
  \caption {Benchmarks from \stranger\ with \texttt{str.replaceall}.\vspace*{-4mm}}
  \label{tab:cvc4-replaceall}
  \begin{tabular}{| l | r | r | r | r | r |} \hline
    Program & \#sat (sec) & \#unsat (sec) & \#to/mo & \#win +/- \\ \hline
    \strsolver-1 & 101 (1,404) & 13 (18) & 6/0 & 9/1 \\ \hline
    \strsolver-2 & 104 (1,178) & 13 (18) & 3/0 & 8/5 \\ \hline
    \strsolver-3 & 103 (772) & 13 (19) & 4/0 & 10/1 \\ \hline
    \strsolver-4 & 101 (316) & 13 (23) & 6/0 & 24/2 \\ \hline
    \strsolver-5 & 102 (520) & 13 (20) & 5/0 & 52/4 \\ \hline
    S3P & 86 (11) & 6 (26) & 0/5 & -- \\ \hline
    \end{tabular}
\vspace*{-4mm}
\end{center}
\end{wraptable}

Our final set of benchmarks is obtained from the third one by filtering out the 120 examples containing \texttt{str.replace} and 
replacing the \texttt{str.replace} operations by \texttt{str.replaceall}, which reflects the real semantics of the original
programs.
This makes the benchmarks more challenging, although they are still simple compared to those of Tables \ref{tab:stranger:php} and \ref{tab:popl}.
The results are shown in Table~\ref{tab:cvc4-replaceall}. The meaning of the
columns is the same as in the previous tables. We compare the different versions
of \strsolver\ against \sthreep\ only since \cvc\ does not support
\texttt{str.replaceall}. On the examples, \sthreep\ crashed 6 times and provided
6 times the unknown result and 13 times a wrong result. 
%
Overall, although \strsolver\ is still slower, it is more reliable than \sthreep\ 
(roughly 10\,\% of wrong and 10\,\% of inconclusive results for \sthreep\ versus
0\,\% of wrong and 5\,\% of inconclusive results for \strsolver).
%

As a final remark, we note that, apart from
experimenting with the \strsolver-1--5 versions, we also tried a version
obtained from \strsolver-3 by replacing the intensionally minimal successor
computation of Section~\ref{sec:afaSolving-min} by the deterministic successor
computation of Section~\ref{sec:afaSolving-det}. On the given benchmark, this
version provided 3 times the best result. This underlines the fact that all of
the described optimizations can be useful in some cases.


%
%
%
%


\section{Conclusions}
\label{sec:conc}

We have presented the first practical algorithm for solving string
constraints with concatenation, general transduction, and regular
constraints; the algorithm is at the same time a decision procedure
for the acyclic fragment $\AC$ of intersection of rational relations 
of \cite{BFL13} and the straight-line fragment $\SL$ of \cite{LB16}. 
The algorithm uses novel ideas including alternating finite automata as
symbolic representations and the use of fast model checkers like IC3
\cite{Bra12} for solving emptiness of alternating automata. In 
initial experiments, our
solver has shown to compare favourably with existing string solvers,
both in terms of expressiveness and performance. More importantly, our solver
can solve benchmarking examples that cannot be handled by existing solvers.

There are several
avenues planned for future work, including more general integration of
length constraints and support for practically relevant operations
like splitting at delimiters and \texttt{indexOf}. Extending our approach to
incorporate a more
general class of length constraints (e.g. Presburger-expressible constraints)
seems to be rather challenging since this possibly would require us to extend
our notion of alternating finite automata with \emph{monotonic counters}
(see \cite{LB16}), which (among others) introduces new problems on how to 
solve language emptiness.

\begin{acks}                            

    Hol\'{i}k and 
Jank{\r u} were supported by the 
\grantsponsor{Czech Science Foundation}{Czech Science Foundation}{} 
(%
project 
\grantnum{16-24707Y}{16-24707Y}%
). 
Hol\'{i}k, Jank{\r u}, and Vojnar were supported by
\grantsponsor{BUT IGA}{the internal BUT grant agency}{} (project \grantnum{FIT-S-17-4014}{FIT-S-17-4014}) 
and the \grantsponsor{IT4IXS: IT4Innovations Excellence in Science}{IT4IXS: IT4Innovations Excellence in Science}{} (project \grantnum{LQ1602}{LQ1602}).
    Lin was supported by \grantsponsor{European Research Council (ERC) under the European
    Union's Horizon 2020 research and innovation programme}{European Research Council (ERC) under the European
    Union's Horizon 2020 research and innovation programme}{} (Grant Agreement
    no \grantnum{759969}{759969}).
    R\"ummer was supported by the
    \grantsponsor{Swedish Research Council}{Swedish Research Council}{}
    under grant \grantnum{2014-5484}{2014-5484}.
%
\end{acks}

\bibliography{references}

\shortlong{}{
\appendix
\begin{center}
    \LARGE \bfseries Appendix
\end{center}

\section{Proof of Lemma~\ref{lem:splittingTerm}}

\begin{proof} Given an $\SL$ conjunction of mixed constraints written in the way
$$ \varphi: \bigwedge_{i=1}^n x_i  =\aftrel_i[\vect s^i](y_1^i\concat\cdots
\concat y_{m_i}^i) $$ 
which satisfies the condition in the definition of $\SL$, we define its weight
as the sum of weights of its variables
$$ \weight \varphi = \sum_{x\in \Var\varphi} \weight x $$
where the weight of a variable $x_i,1\leq i \leq n$, is defined as 
$$ \weight{x_i} =  (m_i-1) + \sum_{j = 1}^{m_i}\weight {y_j^i}, $$
and the weight of a variable $x \in \Var\varphi\setminus\{x_1,\ldots,x_n\}$ as
$\weight x = 0$.
That is, the weight of $x_i$ is derived from its defining constraint as the sum
of the weights of the concatenated variables and the number of concatenation
operators used.

Let the binary splitting step replace $x_i = \aftrel(y_1\concat \cdots \concat
y_m \concat z_1  \concat \cdots \concat z_l)[\vect s^i]$ in $\varphi$ by $x'_1 =
\aftrel_1(y_1 \concat \cdots \concat y_m)[\vect s^i , \vect t] \land x'_2 =
\aftrel_{2}(z_1\concat\cdots\concat z_l)[\vect s^i , \vect t]$, while replacing
every other occurrence of $x_i$ by $x'_1\concat x'_2$, producing the formula
$\varphi'$.
Let us analyse how the splitting influenced the weights. 
First, observe that the weights of variables $x_k$ for $k < i$ and of the
undefined variables outside $\{x_1,\ldots,x_n\}$ do not change since they do not
depend on $x_i$, by the definition of $\SL$.
Hence, the weights of all variables $y_1,\ldots,y_m$ and $z_1,\ldots,z_l$ that
define $\weight{x_i}$ do not change.
Since $x_i$ is in $\Var{\varphi'}$ replaced by $x_1'$ and $x_2'$, the weight
$\weight{x_i}$ of $x_i$ from the sum defining $\weight{\varphi}$ is in the sum
defining $\weight \varphi'$ replaced by $\weight{x_1'}+\weight{x_2'}$.
Observe that due to the $-1$ factor in the definition of a~weight of a~variable,
the second number is smaller by one: 
\begin{multline*}
\weight {x_i} = (m+l-1) + \sum_{j=1}^m \weight {y_j} +  \sum_{j=1}^l \weight
{z_j}  
> \\ 
> 
\Bigl((m-1) + \sum_{j=1}^m \weight {y_j}\Bigr) +  \Bigl((l-1) + \sum_{j=1}^l
\weight {z_j} \Bigr) = \weight{x'_1}+\weight{x'_2}
\end{multline*}

In the defining constraints of variables $x_k$, $k > i$, $x_i$ is replaced by
$x_1' \concat x_2'$; hence, in the sum defining $\weight{x_k}$, $\weight{x_i}$
is replaced by $\weight{x_1'}+\weight{x_2'}+1$ (where the $+1$ is for the one
occurrence of concatenation) which is, by the previous analysis, equal to
$\weight{x_i}$.
The weights of variables $x_k$, $k>i$, in $\varphi'$ are therefore the same as
in $\varphi$.
Since weights of all variables in $\Var\varphi$ are the same for both formulae,
the only difference between the formulae weights is that between $\weight{x_i}$
in $\varphi$ and $\weight{x_1'}+\weight{x_2'}$ in $\varphi'$.
It must therefore hold that $\weight {\varphi} > \weight {\varphi'}$. 

Additionally, observe that $\weight \varphi$ can obviously never be smaller than
$0$ since $\weight x \geq 0$ for every $x\in\Var\varphi$.
This can be shown by an easy induction on $n$.
That concludes the proof.\end{proof}

}

\end{document}